\documentclass[onecolumn,amsmath,showpacs,amssymb]{revtex4}

\usepackage{color}
\usepackage{graphicx}
\usepackage{dcolumn}
\usepackage{bm}
\usepackage{pifont}

\newcommand{\be}[1]{\begin{equation}\label{eq:#1}}
\newcommand{\ee}{\end{equation}}
\newcommand{\bea}{\begin{eqnarray}}
\newcommand{\eea}{\end{eqnarray}}
\newcommand{\bt}{\textbf}

\newcommand{\phd}{\phantom{\dag}}
\newcommand{\ph}{\phantom{.}}

\newcommand{\up}{^{\phd}}
\newcommand{\noi}{\noindent}
\newcommand{\no}{\nonumber}

\newcommand{\RNum}[1]{\bt{{\rm \uppercase\expandafter{\romannumeral #1\relax}}}}

\newenvironment{changemargin}[2]{\begin{list}{}{\setlength{\topsep}{0pt}  \setlength{\leftmargin}{#1}  \setlength{\rightmargin}{#2}
\setlength{\listparindent}{\parindent}
\setlength{\itemindent}{\parindent}  \setlength{\parsep}{\parskip}
}\item[]}{\end{list}}

\newcommand{\lcm}{\begin{changemargin}{-1.3cm}{0.5cm}}
\newcommand{\ecm}{\end{changemargin}}
\newcommand{\tbcm}{\begin{changemargin}{-0.8cm}{-0.25cm}}
\newcommand{\tecm}{\end{changemargin}}

\begin{document}
\def\v#1{{\bf #1}}

\title{Classification of engineered topological superconductors}

\author{Panagiotis Kotetes}\email{panagiotis.kotetes@kit.edu}
\affiliation{Institut f\"{u}r Theoretische Festk\"{o}rperphysik, Karlsruhe Institute of Technology, 76128 Karlsruhe, Germany}

\vskip 1cm
\begin{abstract}
I perform a complete classification of 2d, quasi-1d and 1d topological superconductors which ori\-gi\-nate from the suitable combination of inhomogeneous Rashba spin-orbit
coupling, magnetism and superconductivity. My analysis reveals alternative types of topological superconducting platforms for which Majorana fermions are accessible.
Specifically, I observe that for quasi-1d systems with Rashba spin-orbit coupling and time-reversal violating superconductivity, {\color{black}as for instance due to a finite
Josephson current flow,} Majorana fermions {\color{black}can emerge} \textit{even in the absence} of magnetism. Furthermore, for the classification I also consider situations
where additional ``hidden'' symmetries emerge, with a significant impact on the topological properties of the system. The latter, generally originate from a combination of
space group and complex conjugation operations that separately do not leave the Hamiltonian invariant. Finally, I suggest alternative directions in topological quantum
computing for systems with additional unitary symmetries.
\end{abstract}


\maketitle

\section{Introduction}

\noi The breakthrough concept of emergent Majorana fermions (MFs) in artificial topological {\color{black}superconducting devices}, pioneered by Fu and Kane \cite{Fu and Kane},
motivated a number of recent experiments \cite{MF experiments,Analytis} that have already provided the first promising results. The two authors demonstrated that the helical
surface states of a three-dimensional topological insulator \cite{TI reviews}, with proximity induced superconducting gap $\Delta$, behave as a time-reversal (${\cal T}$)
invariant {\color{black}topological superconductor (TSC)}. When a magnetic field is applied perpendicular to the topological surface, a single MF appears per superconducting
vortex. {\color{black}In fact, the latter mechanism had been discussed earlier by Sato \cite{Sato} in the context of axion-strings.} Shortly after Fu-Kane proposal, it was
recognized that the catalytic presence of spin-momentum locking could be alternatively provided by spin-orbit interaction, {\color{black}intrinsic to non-centrosymmetric
superconductors} \cite{NCS} and Rashba semiconductors \cite{SauSemi,AliceaSemi,Sau,Oreg}. In the case of a semiconducting wire \cite{Sau,Oreg}, fabricated for instance by
InSb, a Zeeman energy $\mu_s|\bm{{\cal B}}|$ is sufficient to lead to MFs localized at the edges. The existence of confined and protected edge MFs is crucial for applications
in topological quantum computing \cite{TQC,Alicea}. A pair of MFs defines a topological qubit, which is in principle \cite{MFdeco} free from decoherence and protected against
noise, in stark contrast to traditional spin \cite{spin qubits} and superconducting qubits \cite{SC qubits}. Furthermore, edge MFs can also give rise to unique transport
signatures \cite{Kitaev,Charge 4pi Josephson,Spin 4pi Josephson, MF transport general}, such as the usual \cite{Kitaev,Charge 4pi Josephson} or the magnetically controlled
\cite{Spin 4pi Josephson} 4$\pi$-Josephson effect.

In the case of a semiconducting wire \cite{Sau,Oreg} with proximity induced superconductivity, the system transits to the topological phase when the criterion $\mu_s|\bm{{\cal
B}}|>\sqrt{\mu^2+|\Delta|^2}$ is satisfied ($\mu$ defines the chemical potential). {\color{black}The concomitant requirement of a high Zeeman energy, which also arises in
quasi-1d multi-channel \cite{multi-channel,Tewari and Sau} analogs of Ref.~\cite{Sau,Oreg}}, can impede the nanofabrication of the device or restrict the possible
mani\-pu\-la\-tions on the MFs. In fact, several proposals concerning quantum information processes rely on the application of strong antiparallel magnetic fields at the
nanoscale level \cite{Flensberg}, something not easily realizable in the lab. As an answer to these obstacles alternative types of engineered TSCs have been put forward, which
support MFs without the necessary presence of spin-orbit coupling or the application of a magnetic field {\color{black}\cite{Alternative MF platforms,PDW,Choy,Flensberg
spiral,Martin,Yazdani}}. In most of these proposals, some kind of inhomogeneous magnetic order coexists with intrinsic or proximity induced superconductivity. For some of
these models \cite{Flensberg spiral,Martin} it has been {\color{black}shown} that there is a mapping to the case of the semiconducting wire-based TSC mentioned above.

In this manuscript I present a complete topological classification of low-dimensional TSCs that support MFs and originate from the combined presence of inhomogeneous 
Rashba spin-orbit coupling $v(\bm{r})$, magnetism $\bm{M}(\bm{r})$ and superconductivity $\Delta(\bm{r})$. My primary goal is to shed light on the topological connection
between different existing proposals for engineered TSCs and in addition to propose alternative advantageous platforms. For my analysis I will consider 2d, quasi-1d and
1d systems. The quasi-1d case is obtained from the strict 2d case by the inclusion of a confining potential $V(\bm{r})$. My study provides new engineered {\color{black}TSCs}
that are experimentally accessible. {\color{black}Specifically, I demonstrate} that for a heterostructure consisting of two coupled single channel Rashba semiconducting wires
deposited on top of a Josephson junction fabricated by two conventional superconductors, MFs can emerge even in the absence of magnetic fields or any type of inhomogeneous
magnetism. In addition, for the classification I examine the effects of dimensionality on the robustness of MFs through separating the systems under investigation into weak
and strong engineered TSCs. Furthermore, I illustrate that so far overlooked discrete symmetries, that I shall refer to as ``hidden" symetries (${\cal O}$), distinguish models
previously considered as topologically equivalent. Generally, hidden symmetries can be either unitary or anti-unitary and result from a combination of space group,
time-reversal or other internal symmetry operations that when considered separately do not leave the Hamiltonian invariant (e.g. \cite{SatoHidden,Hidden AFM,Hidden SU(4)}).
Here I discuss two examples of hidden symmetries: i) a unitary hidden symmetry resulting from the combination of a reflection and a translation and ii) an anti-unitary
symmetry resulting from the combination of time-reversal and translation operations. Finally, I also discuss new topological quantum computing (TQC) perspectives that appear
when additional {\color{black}unitary symmetries, including hidden symmetries}, are present.   

At this point, I give a brief description of how the several sections are organized. In Section~{\color{black}\RNum{2}}, I provide a short introduction to Majorana fermions and
introduce the general Hamiltonian that describes the systems of interest. In Section~{\color{black}\RNum{3}}, I shortly review the topological classification methods with
special focus on the situations where additional unitary and anti-unitary symmetries are present. In Section~{\color{black}\RNum{4}}, I present an overview of my main results
(Table~\ref{table:Topological Phases}) concerning the classification of TSCs when all possible spatial symmetries are broken. I further discuss how the emergence of hidden
symmetries can modify Table~\ref{table:Topological Phases}. In Section~{\color{black}\RNum{5}}, I provide a detailed analysis and justification of the results presented in
Section~{\color{black}\RNum{4}}. {\color{black}In Section~{\color{black}\RNum{6}}, I demonstrate that MFs are accessible in heterostructures consisting of conventional
superconductors in proximity to \bt{A.} the surface states of a 3d topological insulator or \bt{B.} two coupled single channel Rashba semiconducting wires, when in both cases
a Josephson current is injected to the system. In Section~{\color{black}\RNum{7}}, I present two specific examples of systems characterized by a hidden symmetry and study the
impact of the latter on the topological properties. In Section~{\color{black}\RNum{8}}, I discuss how the presence of hidden symmetries can be useful for developing topological
quantum computing protocols and suggest possible candidate systems that could be used for this purpose. Finally, Section~{\color{black}\RNum{9}} summarizes my main results and
related conclusions.

\section{Majorana fermions and model Hamiltonian}

\noi In condensed matter physics MFs are not fundamental particles \cite{Majorana himself} but excitations of a many-body system \cite{Wilczek,Read}. Essentially, what we
define as MFs are the operators $\gamma_{\alpha}$ ($\alpha$ is just a label) which satisfy $\{\gamma_{\alpha},\gamma_{\beta}\}=\delta_{\alpha,\beta}{\rm I}$ (${\rm I}$ the
identity operator) and constitute zero energy eigen-operators of the Bogoliubov - de Gennes (BdG) Hamiltonian. Since MFs are hermitian they can be described by the following
general expression
\bea
\gamma_{\alpha}=\int d\bm{r}\left[u_{\uparrow,\alpha}^*(\bm{r})\psi_{\uparrow}(\bm{r})+u_{\downarrow,\alpha}^*(\bm{r})\psi_{\downarrow}(\bm{r})+
u_{\uparrow,\alpha}(\bm{r})\psi_{\uparrow}^{\dag}(\bm{r})+u_{\downarrow,\alpha}(\bm{r})\psi_{\downarrow}^{\dag}(\bm{r})\right]\,,
\eea

where $\psi_{\sigma}^{\dag}(\bm{r})$/$\psi_{\sigma}\up(\bm{r})$ correspond to the creation/annihilation operators of an electron {\color{black}with position vector $\bm{r}$
(here $\bm{r}=(x,y)$) and} spin projection $\sigma=\uparrow,\downarrow$. Notice that MFs require linear combinations of electronic operators and their hermitian conjugates.
{\color{black}Consequently, in order for MFs to constitute the only type of accessible eigen-operators of the single particle Hamiltonian, we have to restrict ourselves} to
systems in which the spin-quantization axis is fixed. Notice that for a system with spin-rotational symmetry, the application of a homogeneous magnetic field breaks the latter
symmetry but the spin-quantization axis can be always redefined. In this case, MFs are not {\color{black}accessible \textit{directly} but only as constituent operators
of electronic eigen-operators}. As a matter of fact, MFs can {\color{black}fundamentally} appear only in systems with {\color{black} spin-orbit coupling, spin-triplet
superconductivity or magnetism with spatially dependent polarization}.

In this work I focus on systems that satisfy the above requirements and are either microscopically or phenomenologically (for heterostructures) described by the following
Hamiltonian 
\bea
{\cal H}&=&\int d\bm{r}\ph\hat{\psi}^{\dag}(\bm{r})\left[\frac{\hat{\bm{p}}^2}{2m}-\mu+V(\bm{r})-\bm{M}(\bm{r})\cdot\bm{\sigma}+
\frac{\{v(\bm{r}),\hat{p}_x\sigma_y-\hat{p}_y\sigma_x\}}{2}\right]\hat{\psi}(\bm{r})\no\\
&+&\int d\bm{r}
\left[\psi_{\uparrow}^{\dag}(\bm{r})\Delta(\bm{r})\psi_{\downarrow}^{\dag}(\bm{r})+\psi_{\downarrow}\up(\bm{r})\Delta^*(\bm{r})\psi_{\uparrow}\up(\bm{r})\right]\,,
\label{eq:Hamiltonian}
\eea

where $\hat{\psi}^{\dag}(\bm{r})=(\psi_{\uparrow}^{\dag}(\bm{r})\phd\psi_{\downarrow}^{\dag}(\bm{r}))$, $\bm{\sigma}$ are the spin Pauli matrices, $v(\bm{r})$ is
the spatially dependent strength of the Rashba spin-orbit coupling, $\bm{M}(\bm{r})$ corresponds to a magnetic-field or a magnetization profile and $\Delta(\bm{r})$ defines 
a spatially varying superconducting order parameter. Notice that in some sense the above Hamiltonian is overcomplete, since it covers all the cases that we will consider,
without implying that all the terms are simultaneously required for obtaining a TSC. Furthermore, at the level of my topological classification, the origin of the involved
terms is unimportant. However, I have to remark that when I will discuss specific cases I will concentrate on engineered {\color{black}TSCs}, which for instance involve
conventional types of magnetism and mainly proximity induced superconductivity \cite{Proximity effect}. This {\color{black}implies} that I will not consider here the cases of
unconventional \cite{commentU} density waves \cite{UDW,Raghu TDW,Chirality Nernst, Topological Meissner,Tewari PKE} or superconductors \cite{MFHighTc}, although some of
{\color{black} the} conclusions could be also applied to these systems.

Since for the situations considered in the present study the spin-quantization is always fixed, I will employ the following spinor 
\bea
\widehat{\Psi}^{\dag}(\bm{r})=\left(\psi_{\uparrow}^{\dag}(\bm{r})\,,\psi_{\downarrow}^{\dag}(\bm{r})\,, \psi_{\uparrow}\up(\bm{r})\,,\psi_{\downarrow}\up(\bm{r})\right)\,,
\eea

and use the $\bm{\tau}$ Pauli matrices in order to represent matrices in the Nambu particle-hole space. With the introduction of the above enlarged spinor the 
Hamiltonian can be rewritten in the following compact way
\bea
{\cal H}=\frac{1}{2}\int d\bm{r}\ph\widehat{\Psi}^{\dag}(\bm{r})\widehat{{\cal H}}(\hat{\bm{p}},\bm{r})\widehat{\Psi}(\bm{r})\,,
\eea

where $\widehat{{\cal H}}(\hat{\bm{p}},\bm{r})$ corresponds to the BdG Hamiltonian. Notice that the factor of $1/2$ is crucial for avoiding double counting of the degrees
of freedom, since the  above spinor does not obey to the usual fermionic commutation relations.

\section{Topological classification principles}

\noi Before discussing the possible topological phases arising from our model Hamiltonian, I will briefly review the basics of how to classify topological systems. My 
goal is to first highlight a key point which is crucial for classifying TSCs and then demonstrate how this can provide further topological insight concerning previously
studied systems \cite{Flensberg spiral, Martin}. This key point is that topological classification of systems following the recently developed methods
{\color{black}\cite{AZ,Tenfold,Kitaev Periodic Table}}, is conducted for irreducible Hamiltonians, for which one cannot find any unitary operator ${\cal O}_u$ satisfying
$[\widehat{{\cal H}}(\hat{\bm{p}},\bm{r}),{\cal O}_u]=0$. If there is a number of these type of operators, we can block diagonalize the Hamiltonian and topologically classify
each sub-block. Of course, this is not the only route to study topological properties, since one can also directly construct topological invariants for reducible Hamiltonians
\cite{Volovik book}. Nevertheless, studying irreducible Hamiltonians provides a transparent analysis of the topological classes. 

The symmetry class and the related accessible topological phases of an irreducible Hamiltonian are defined by the possible presence of three specific types of discrete
symmetries. The first two correspond to a generalized time-reversal symmetry effected by the anti-unitary operator $\Theta$ and a charge conjugation symmetry effected by an
anti-unitary operator $\Xi$. If $\Theta$ is a symmetry of the Hamiltonian, it satisfies $[\widehat{{\cal H}}(\hat{\bm{p}},\bm{r}),\Theta]=0\Rightarrow\Theta^{-1}\widehat{{\cal
H}}(\hat{\bm{p}},\bm{r})\Theta=+\widehat{{\cal H}}(\hat{\bm{p}},\bm{r})$ while in the case of charge-conjugation we instead have $\{\widehat{{\cal
H}}(\hat{\bm{p}},\bm{r}),\Xi\}=0\Rightarrow\Xi^{-1}\widehat{{\cal H}}(\hat{\bm{p}},\bm{r})\Xi=-\widehat{{\cal H}}(\hat{\bm{p}},\bm{r})$. If $\Theta$ and $\Xi$ constitute
symmetries of the Hamiltonian at the same time, then the Hamiltonian additionaly satisfies $\{\widehat{{\cal H}}(\hat{\bm{p}},\bm{r}),\Theta\Xi\}=0$ where the combined
$\Theta\Xi$ operator is unitary and is termed chiral symmetry operator $\Pi$. The inclusion of $\Pi$ completes the set of symmetries that are required for determining the
symmetry class of an irreducible Hamiltonian. In fact, in order to cover all possible symmetry classes, we have to take into account the case in which a unitary chiral
symmetry may exist without the necessary presence of $\Theta$ and $\Xi$ symmetries.    

{\color{black}Another important aspect which has not been pointed out so far in the existing classification schemes, concerns correlated systems and the role of induced order
parameters \cite{CMR,patterns,Ce,GV} on the topological properties of a system. Within a mean-field description, it has been shown that there exist patterns \cite{patterns,GV}
of thermodynamic phases and their corresponding order parameters, which are bound to coexist at a microscopic level. In fact, in Ref~\cite{GV}, Varelogiannis recently put
forward a rule according to which one can predict the induced order parameters and consequently the complete patterns of thermodynamic/topological phases which can be
decomposed in fundamental coexistence quartets of phases. Although the symmetry properties of an induced order parameter is strictly determined by the already existing order
parameters and consequently cannot alter the symmetry class, its inclusion can deform the topological phase diagram by modifying the parameter regime for observing the
accessible topological classes.} 

In the present work I am interested in ``hidden'' unitary discrete symmetry operators satisfying the property ${\cal O}_u^n={\rm I}$, with $n\in\mathbb{Z}$. In the simplest
case $n=2$, we can block diagonalize the Hamiltonian into two sub-blocks labelled by the eigenvalues $\pm 1$ of ${\cal O}_u$, leading to a direct sum of the form
$\widehat{{\cal H}}_{+}(\hat{\bm{p}},\bm{r})\oplus\widehat{{\cal H}}_{-}(\hat{\bm{p}},\bm{r})$. Notice that because of the discrete symmetry ${\cal O}_u$, both sub-blocks are
constrained to belong to the same symmetry class. However, the two sub-systems do not necessarily reside in the same topological class. In addition, I also provide an example
of an anti-unitary hidden symmetry ${\cal O}_a$. In this case ${\cal O}_a$ constitutes an additional generalized time-reversal symmetry which modifies the initial symmetry
class of the system, instead of splitting the latter in a direct sum of identical symmetry classes as for the unitary analog ${\cal O}_u$. For instance, if a system is
initially in class D, then the emergence of an anti-unitary hidden symmetry ${\cal O}_a$ with ${\cal O}^2_a=+{\rm I}$ will change its {\color{black}symmetry to class BDI}.

For the cases under consideration, the BdG Hamiltonian enjoys a charge-conjugation symmetry $\Xi=\tau_x{\cal K}$, where ${\cal K}$ defines complex conjugation. Since
$\Xi^2=+{\rm I}$, we obtain only the following three allowed symmetry classes presented in Table \ref{table:Classes}: BDI, D, DIII or their direct sums BDI$\oplus$BDI,
D$\oplus$D, DIII$\oplus$DIII in the presence of a hidden symmetry ${\cal O}_u$, with ${\cal O}_u^2={\rm I}$. Notice that the classes BDI and DIII are characterized by a
time-reversal symmetry $\Theta$ with $\Theta^2=+{\rm I}$ and $\Theta^2=-{\rm I}$, respectively. In the first case, $\Theta$ symmetry implies that the Hamiltonian is real while
in the second that there exist a Kramers-type degeneracy leading to doublets of solutions. Below I examine the minimal cases that can lead to a symmetry class supporting MFs.
For completeness I will also shortly discuss previously studied models.

\begin{table}[h]\caption{Symmetry classes of topological superconductors supporting Majorana fermions {\color{black}\textit{``fundamentally''}, i.e. the eigenope\-rators
diagonalizing the single-particle Hamiltonian are solely of the Majorana type}. For $\Theta$ and $\Xi$, ${\rm \pm I}$ corresponds to the result of $\Theta^2$ and $\Xi^2$. For
$\Pi$, ${\rm I}$ denotes that the symmetry is present. Conversely, $0$ implies that the corresponding symmetry is broken.}\vspace{0.1in}
\begin{tabular}{|c|c|c|c|c|c|c|}\hline
{\rm \bt{Class}}&$\Theta$&$\Xi$&$\Pi$&{\rm 1d}&{\rm 2d}&{\rm 3d}\\\hline
{\rm BDI}&${\rm+I}$&${\rm+I}$&${\rm I}$&$\bm{\mathbb{Z}}$&$0$&$0$\\\hline
{\rm D}&$0$&${\rm+I}$&$0$&$\bm{\mathbb{Z}}_2$&$\bm{\mathbb{Z}}$&$0$\\\hline
{\rm DIII}&${\rm-I}$&${\rm+I}$&${\rm I}$&$\bm{\mathbb{Z}}_2$&$\bm{\mathbb{Z}}_2$&$\bm{\mathbb{Z}}$\\\hline
\end{tabular}
\label{table:Classes}
\end{table}

{\color{black}
\section{Results: allowed phases of engineered topological superconductors}
}
\noi In the present section I carry out a thorough analysis of the accessible TSC phases that follow from the Hamiltonian of Eq.~(\ref{eq:Hamiltonian}). For the strict 2d
and 1d cases I will consider that $V(\bm{r})=0$. To analyze the quasi-1d case, I will always assume the presence of a confining potential $V(y)$. For topological
computation applications based on edge MFs the quasi-1d and pure 1d setups are the most relevant. The possible unitary symmetries that can appear for these systems originate
from the point group ${\cal G}$ and translation operations $t_{\bm{a}}$ with $t_{\bm{a}}\bm{r}=\bm{r}+\bm{a}$. Let me now focus on the point group symmetries for the quasi-1d
and pure 1d geometries, which I depict in Fig.~\ref{fig:Symmetries}. The point group for a quasi-1d system confined in the $xy$-plane is C$_{2v}$. This
symmetry group includes a C$_2$ $\pi$-rotation about the $z$-axis ($\bm{r}\rightarrow-\bm{r}$, $z\rightarrow z$) and two $\sigma_v$ reflection operations $\sigma_{xz}$
($y\rightarrow-y$) and $\sigma_{yz}$ ($x\rightarrow-x$), where the indices correspond to the mirroring plane. Notice that the reflection symmetry operation
$\sigma_h\equiv\sigma_{xy}$ ($z\rightarrow-z$) is broken in C$_{2v}$. In the strict 1d case we are left only with $\sigma_{yz}$. For random $v(\bm{r})$, $\bm{M}(\bm{r})$ and
$\Delta(\bm{r})$ all the aforementioned symmetries are broken. Nevertheless, for special spatial profiles of the latter functions, a hidden symmetry can emerge, which
consists of these basic symmetry operations or other \textit{already broken} symmetries such as $\sigma_h$.

\begin{figure}[b]
\includegraphics[width=3.2in,height=1.9in]{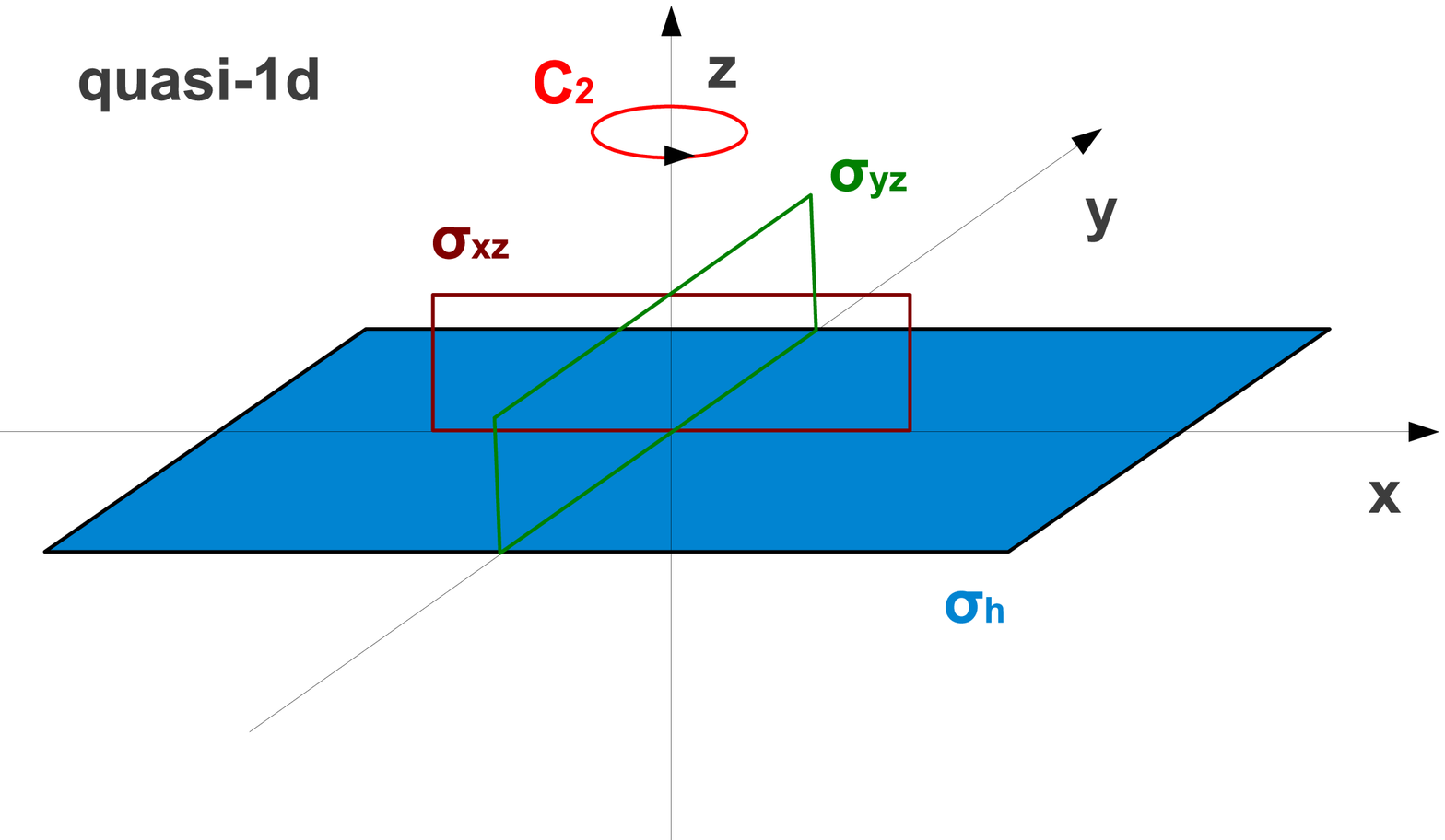}\hspace{0.4in}
\includegraphics[width=3.2in,height=1.9in]{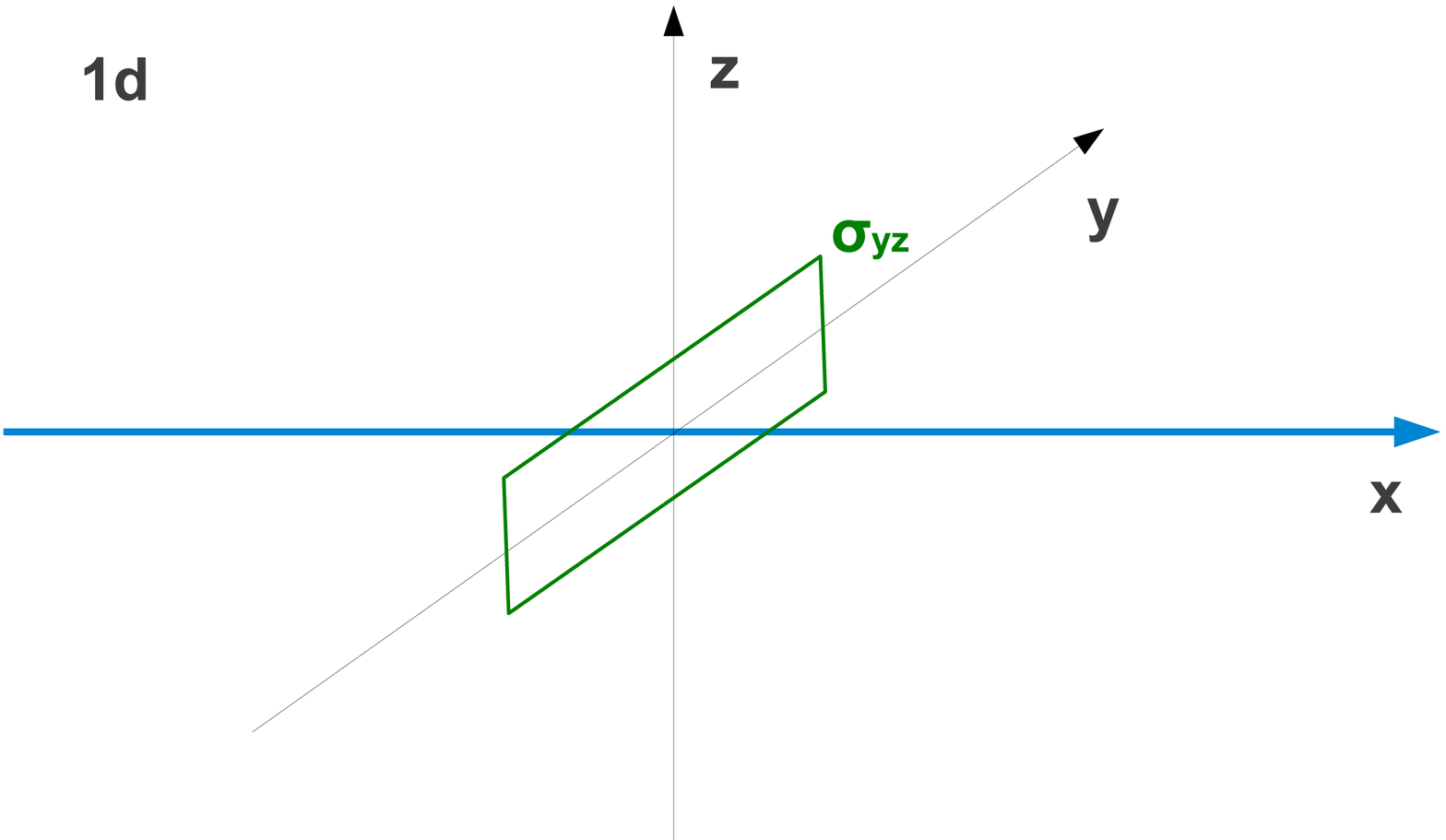}
\caption{Point group symmetries for quasi-1d and strictly 1d geometries of a topological superconductor (depicted with blue). In the quasi-1d case, inversion symmetry along
the $z-$axis, $\sigma_h$, is broken.}
\label{fig:Symmetries}
\end{figure}

In Table~\ref{table:Topological Phases}, I present the topological classification for the Hamiltonian of Eq.~(\ref{eq:Hamiltonian}) where all possible unitary symmetries
are broken due to the spatial dependence of $v(\bm{r})$, $\bm{M}(\bm{r})$ and $\Delta(\bm{r})$. I demand that $|\bm{M}(\bm{r})|\neq0$ and $|\Delta(\bm{r})|\neq0$ so as to
avoid any gap closings that could lead to a macroscopic coexistence of different topological phases throughout the volume of the material. I also have to remark that in the
case of a translationally invariant system, we can transfer to $\bm{k}-$space in order to calculate topological invariants. If translational symmetry is broken, then analysis
of the topological properties in coordinate or momentum space exhibits the same complexity. Of course, there can be also cases where topological properties in combined
$(\bm{r},\bm{k})$-space can be relevant \cite{Volovik book,ZhangTQFT}.

\begin{table}[t]\caption{Accessible topological superconducting phases {\color{black}supporting MFs} due to the combined presence of inhomogeneous Rashba spin-orbit coupling
$v(\bm{r})$, magnetization $\bm{M}(\bm{r})$ and superconducting order parameter $\Delta(\bm{r})$. The presence (absence) of the aforementioned terms is indicated with
{\color{black}\checkmark} ({\color{black}\ding{55}}). The resulting symmetry class depends on the behaviour of the magnetic and superconducting Hamiltonian terms under complex
conjugation ${\cal K}$, since the Rashba spin-orbit coupling term always preserves ${\cal T}$. For a term that is already present ({\color{black}\checkmark}), we denote the case
of preserved (broken) complex conjugation as ${\cal K}={\rm I}~(0)$. Notice that symmetry classes that lack a strong topological invariant for the corresponding dimensionality
are shown with italics and correspond to weak topological superconductors. With the phrase {\color{black}``\textit{no MFs}''}, I imply that the system belongs to a symmetry
class, other than D, BDI and DIII, which cannot {\color{black}\textit{fundamentally}} support MFs. In the presence of a unitary hidden symmetry ${\cal O}_u$ with the property
${\cal O}_u^n={\rm I}$, a symmetry class TC splits into $n$ identical sub-classes $\oplus_n$TC. However, the sub-systems do not have to reside in the same topological class.
{\color{black}Note {\color{black} that} the identification of the symmetry class does not necessarily imply that a system can indeed transit to the topologically non-trivial
regime hosting MFs. This depends on the particular implementation.}}
\vspace{0.2in}
\begin{tabular}{|c|c|c|c|c|c|c|}\hline
{\rm \bt{Case}}&$v(\bm{r})$&$\bm{M}(\bm{r})$&$\Delta(\bm{r})$&{\rm 2d}&{\rm quasi-1d}&{\rm 1d}\\\hline
\RNum{1}&{\color{black}\checkmark}&{\color{black}\ding{55}}&${\cal K}={\rm I}$&{\rm\bt{DIII}}&{\rm \bt{DIII}}&\textit{no MFs}\\\hline
\RNum{2}&{\color{black}\checkmark}&{\color{black}\ding{55}}&${\cal K}={\rm 0}$&{\rm \bt{D}}&{\rm \bt{D}}&\textit{no MFs}\\\hline
\RNum{3}&{\color{black}\ding{55}}&${\cal K}={\rm I}$&${\cal K}={\rm I}$&{\color{black}\textit{BDI}}&{\rm \bt{BDI}}&{\rm \bt{BDI}}\\\hline
\RNum{4}&{\color{black}\ding{55}}&${\cal K}={\rm \{0,I,0\}}$&${\cal K}={\rm \{I,0,0\}}$&{\rm \bt{D}}&{\rm \bt{D}}&{\rm \bt{D}}\\\hline
\RNum{5}&{\color{black}\checkmark}&${\cal K}={\rm I}$&${\cal K}={\rm I}$&{\rm \bt{D}}&{\rm \bt{D}}&{\rm \bt{BDI}}\\\hline
\RNum{6}&{\color{black}\checkmark}&${\cal K}={\rm \{0,I,0\}}$&${\cal K}={\rm \{I,0,0\}}$&{\rm \bt{D}}&{\rm \bt{D}}&{\rm \bt{D}}\\\hline
\end{tabular}
\label{table:Topological Phases}
\end{table}

One of the most important results is Case \RNum{2} in Table~\ref{table:Topological Phases}, where the simultaneous presence of Rashba spin-orbit coupling and inhomogeneous
superconductivity can lead to MFs in a quasi-1d system, without the requirement of a magnetic field. In fact, an $\bm{r}$-dependent superconducting phase originating from a
supercurrent falls into this case, constituting an experimentally prominent route towards MFs. As far as the table is concerned, the possible phases are essentially classified
by the behaviour of the magnetic and superconducting Hamiltonian terms under ${\cal K}$.\\

\section{Analysis of the possible topological phases in the absence of unitary symmetries}

\noi In this section I provide the detailed topological classification for the cases presented in Table \ref{table:Topological Phases}. Notice that for the present discussion
the spatial dependence of the terms involved is considered random, unless explicitly stated.

\begin{itemize}

\item \bt{Cases \RNum{1} and \RNum{2}}

In the following paragraph I will focus on the Cases \RNum{1} and \RNum{2} that are characterized by the presence of inhomogeneous Rashba spin-orbit coupling $v(\bm{r})$ and
superconducting order parameter $\Delta(\bm{r})$. The TSCs belonging to these cases are described by the following Hamiltonian
\bea
\widehat{{\cal H}}(\hat{\bm{p}},\bm{r})=
\left[\frac{\hat{\bm{p}}^2}{2m}-\mu+V(\bm{r})\right]\tau_z+\frac{\left\{v(\bm{r}),\hat{p}_x\tau_z\sigma_y-\hat{p}_y\sigma_x\right\}}{2}-\Delta_{\Re}(\bm{r})\tau_y\sigma_y-
\Delta_{\Im}(\bm{r})\tau_x\sigma_y\,.
\label{eq:Hamiltonian Rashba and superconductivity}
\eea

The Rashba spin-orbit coupling term is odd under inversion symmetry along the $z$-axis $\sigma_h$, while it is even under the usual time-reversal symmetry ${\cal T}$. If the
superconducting term is also invariant under ${\cal T}$ or equivalently ${\cal K}$, since we are dealing with a scalar superconducting order parameter, then
$\Delta(\bm{r})=\Delta_{\Re}(\bm{r})$ and the full Hamiltonian is characterized by the generalized time-reversal symmetry $\Theta=i\sigma_y{\cal K}$ that coincides with ${\cal
T}$. 

\noi \bt{2d system:} In the 2d case, the particular system belongs to the symmetry class DIII and is related to the model of Ref.~\cite{Fu and Kane}. Since $\Theta$
satisfies $\Theta^2=-{\rm I}$, with ${\rm I}$ the identity operator, we expect boundary MF Kramers doublets. Class DIII possesses a strong $\mathbb{Z}_2$ topological invariant
in 2d. The presence of $\Theta$ also leads to a chiral symmetry with $\Pi=\tau_x\sigma_y$. In the case where the superconducting order parameter has an additional imaginary
component, ${\cal T}$ is broken and the system transits to class D. Class D has a strong $\mathbb{Z}$ invariant in 2d and consequently this system constitutes a strong TSC in
both cases.

In order to analyze the symmetry properties in a more transparent manner, I will consider without any loss of generality, the following form for the superconducting order
parameter $\Delta(\bm{r})=\Delta e^{i\bm{J}\cdot\bm{r}}$. The particular profile, constitutes the simplest representative of ${\cal T}$ violating superconductivity and can be
viewed either as the result of the spontaneous formation of a Fulde-Ferrell \cite{Fulde} phase with modulation wave-vector $\bm{J}$ or the consequence of the application of a
supercurrent $\bm{J}$. The Fulde-Ferrell phase is a special case of pair density waves {\color{black}(see also \cite{PDWgeneral})} that have been also recently considered
\cite{PDW} as potential TSCs leading to MFs. On the other hand, the application of supercurrents was previously discussed in Refs.~\cite{MFsupercurrents}. In the latter
implementations a supercurrent was viewed as an {\color{black}additional} knob for tuning the topological phase diagram, {\color{black}without though being a necessary
ingredient for obtaining a TSC.}

At this point we proceed with gauging away the superconducting phase $\varphi(\bm{r})=\bm{J}\cdot\bm{r}$ via the minimal
coupling $\hat{\bm{p}}\rightarrow\hat{\bm{p}}+\hbar\bm{\nabla}\varphi(\bm{r})\tau_z/2=\hat{\bm{p}}+\hbar\bm{J}\tau_z/2$, leading to 
\bea
\widehat{{\cal H}}'(\hat{\bm{p}}{\color{black},\bm{r}})&=&\frac{\hbar}{2m}\ph\bm{J}\cdot\hat{\bm{p}}\ph I+\left[\frac{\hat{\bm{p}}^2}{2m}+
\frac{\left(\hbar\bm{J}/2\right)^2}{2m}-\mu+V(\bm{r})\right]\tau_z+\frac{\left\{v(\bm{r}),\hat{p}_x\tau_z\sigma_y-\hat{p}_y\sigma_x\right\}}{2}+\frac{v(\bm{r})\hbar}{2}
\left(J_x\sigma_y-J_y\tau_z\sigma_x\right)\no\\&-&\Delta\tau_y\sigma_y\,.\label{eq:Gauged Hamiltonian}
\eea

It is straightforward to confirm that for $\bm{J}=\bm{0}$ the system belongs to class DIII because ${\cal T}$ is preserved while for finite $\bm{J}$ the system lies in 
class D. 

\noi \bt{quasi-1d system:} In order to investigate the quasi-1d and 1d cases I set $v(\bm{r})=v(x)$. Furthermore for the quasi-1d case I additionaly switch on a confining
potential $V(\bm{r})=V(y)$. The presence of the confining potential lowers the symmetry of the system, permitting anisotropic coefficients for the Rashba terms
$\hat{p}_x\tau_z\sigma_y$ and $\hat{p}_y\sigma_x$, instead of a common $v(x)$. For my analysis I will keep the coefficients equal since the only crucial requirement for my
study is that they are both non-zero. To achieve confinement, I consider the case of a harmonic potential $V(y)=m\omega^2y^2/2$. This term is translationally invariant along
the $x$-direction and even under C$_2$, $\sigma_{xz}$ and $\sigma_{yz}$. Another option for the confining potential is the infinite wall potential $V(|y|>L_y)=+\infty$. For
the choice of the harmonic confining potential, the Hamiltonian reads
\bea
\widehat{{\cal H}}(\hat{p}_x{\color{black},x},\hat{a},\hat{a}^{\dag})&=&\frac{\hbar}{2m}\left(J_x\hat{p}_x+J_y\sqrt{\frac{m\omega\hbar}{2}}\ph\frac{\hat{a}-\hat{a}^{\dag}}{i}
\right)I+\left[\frac{\hat{p}^2_x}{2m}+\frac{(\hbar\bm{J}/2)^2}{2m}-\mu+\hbar\omega\left(\hat{a}^{\dag}\hat{a}+\frac{1}{2}\right)\right]\tau_z\no\\&+&
\frac{\{v(x),\hat{p}_x\}}{2}\ph\tau_z\sigma_y-v(x)\sqrt{\frac{m\omega\hbar}{2}}\ph\frac{\hat{a}-\hat{a}^{\dag}}{i}\sigma_x
+\frac{v(x)\hbar}{2}\left(J_x\sigma_y-J_y\tau_z\sigma_x\right)-\Delta\tau_y\sigma_y\,,
\eea
 
where I introduced the quantum harmonic oscillator's bosonic creation (annihilation) operator $\hat{a}^{\dag}$ ($\hat{a}$). By introducing the eigenfunctions
$\left|n\right>$ of the number operator $\widehat{N}=\hat{a}^{\dag}\hat{a}$, I obtain the matrix Hamiltonian
\bea
\widehat{{\cal H}}(\hat{p}_x{\color{black},x})&=&\frac{\hbar}{2m}\left(J_x\hat{p}_xI+J_y\hbar\lambda_y\right)+\left(\frac{\hat{p}^2_x}{2m}-\mu\lambda_z^{\bm{J}}\right)\tau_z+
\frac{\{v(x),\hat{p}_x\}}{2}\ph\tau_z\sigma_y-v(x)\hbar\lambda_y\sigma_x+\frac{v(x)\hbar}{2}\left(J_x\sigma_y-J_y\tau_z\sigma_x\right)\no\\&-&\Delta\tau_y\sigma_y\,,
\label{eq:Hamiltonian quasi-1d random Rashba}
\eea

that is defined in spin, Nambu and $\widehat{N}$ spaces with
\bea
\left<n\right|\lambda_z^{\bm{J}}\left|s\right>=\delta_{n,s}\left[\mu-\frac{(\hbar\bm{J}/2)^2}{2m}-\hbar\omega\left(n+\frac{1}{2}\right)\right]/\mu\quad{\rm and}\quad
\left<n\right|\lambda_y\left|s\right>=\sqrt{\frac{m\omega}{2\hbar}}\ph\frac{\sqrt{n+1}\delta_{n,s-1}-\sqrt{n}\delta_{n,s+1}}{i}\,.
\eea
 
Since the form of the Hamiltonian is identical to the 2d case and ${\cal K}^{-1}\lambda_y{\cal K}=-\lambda_y$ (similarly to $\hat{p}_y$), the quasi-1d model also 
belongs to the DIII class {\color{black}Ref.~\cite{comment}} for $\bm{J}=\bm{0}$ and to class D for $\bm{J}\neq\bm{0}$.

\noi \bt{1d system:} For studying the strictly 1d system, I apply the dimensional reduction method to the 2d model of Eq.~(\ref{eq:Gauged Hamiltonian}) and set
$\hat{p}_y=J_y=0$, that yields
\bea
\widehat{{\cal H}}'(\hat{p}_x{\color{black},x})=\frac{\hbar}{2m}\ph J_x\hat{p}_x I+\left[\frac{\hat{p}_x^2}{2m}+\frac{\left(\hbar J_x/2\right)^2}{2m}-\mu\right]\tau_z+
\frac{\{v(x),\hat{p}_x\}}{2}\ph\tau_z\sigma_y+\frac{v(x)}{2}\ph\hbar J_x\sigma_y-\Delta\tau_y\sigma_y\,.\\\no
\eea

We observe that for this model we retain our freedom to redefine the spin-quantization axis and as a result the above Hamiltonian does not support MFs {\color{black} in a
fundamental manner. If we rotate the spin-quantization axis from $y$ to $z$, we can rewrite the above Hamiltonian using the usual two-component Nambu spinor
$\hat{\psi}^{\dag}_N(x)=(\psi_{\uparrow}^{\dag}(x),\psi_{\downarrow}(x))$, since the four-component formalism becomes redundant in this case. In this formalism the
eigenoperators are electronic and their decomposition into MF operators can serve as an equivalent but not necessary description. For instance, if $J_x=0$, the
Hamiltonian in the latter formalism belongs to class AIII which is characterized by a $\mathbb{Z}$ topological invariant in 1d. In this case, the system can support
zero-energy edge electronic eigenoperators which can be decomposed into edge MFs. In this sense, MFs are not fundamental in the 1d case.}

\item \bt{Cases \RNum{3} and \RNum{4}}

In this section I consider TSC phases that do not involve spin-orbit coupling. This implies that {\color{black}\textit{at least two components}} of an inhomogeneous
magnetization field must be present in order to lock the spin-quantization axis, since the latter constitutes a prerequisite for obtaining MFs. For this kind of systems, 
the Hamiltonian reads
\bea
\widehat{{\cal H}}(\hat{\bm{p}},\bm{r})=\left[\frac{\hat{\bm{p}}^2}{2m}-\mu+V(\bm{r})\right]\tau_z-\bm{M}(\bm{r})\cdot\left(\tau_z\sigma_x,\sigma_y,\tau_z\sigma_z\right)
-\Delta_{\Re}(\bm{r})\tau_y\sigma_y-\Delta_{\Im}(\bm{r})\tau_x\sigma_y\,.
\eea

For the specific type of TSCs, the magnetization field $\bm{M}(\bm{r})$ is odd under the usual time-reversal symmetry ${\cal T}$. However, its behavior under complex
conjugation ${\cal K}$ is not fixed. If $M_y(\bm{r})=0$ then $\bm{M}(\bm{r})$ preserves ${\cal K}$. This leads to the following two possibilities depending also on the
behaviour of the superconducting order parameter under ${\cal K}$. In the first possibility the magnetic and superconducting terms are simultaneously invariant under ${\cal
K}$ and a generalized time-reversal symmetry appears with $\Theta={\cal K}$ accompanied by a chiral symmetry $\Pi=\tau_x$.

\noi \bt{2d system:} In 2d, the system belongs to the BDI class that however is not characterized by a strong topological invariant for this dimensionality. Consequently, the
specific system corresponds to a weak TSC, since under special circumstances one could define weak invariants. The second possibility involves the breaking of ${\cal
K}$ by either one of the terms. In the latter case, the Hamiltonian belongs to class D which has a strong $\mathbb{Z}$ topological invariant in 2d.

\noi \bt{quasi-1d system:} For the particular study I will consider for convenience that $\Delta(\bm{r})=\Delta e^{i\bm{J}\cdot\bm{r}}$. As previously, I gauge away the
superconducting phase and obtain the equivalent model
\bea
\widehat{{\cal H}}'(\hat{\bm{p}}{\color{black},\bm{r}})=\frac{\hbar}{2m}\ph\bm{J}\cdot\hat{\bm{p}}\ph I+\left[\frac{\hat{\bm{p}}^2}{2m}+
\frac{\left(\hbar\bm{J}/2\right)^2}{2m}-\mu+V(\bm{r})\right]\tau_z-\bm{M}(\bm{r})\cdot\left(\tau_z\sigma_x,\sigma_y,\tau_z\sigma_z\right)-\Delta\tau_y\sigma_y\,.
\label{eq:Gauged Hamiltonian magnetism}
\eea

For effecting confinement I will employ once again a harmonic oscillator's potential $V(y)=m\omega^2y^2/2$ and we also have
\mbox{$\bm{M}(\bm{r})=$} \mbox{$\bm{M}(x,\hat{a}+\hat{a}^{\dag})$}. Following
the same steps as in Cases \RNum{1} and \RNum{2}, I obtain the Hamiltonian
\bea
\widehat{{\cal
H}}'(\hat{p}_x{\color{black},x})=\frac{\hbar}{2m}\left(J_x\hat{p}_xI+J_y\hbar\lambda_y\right)+\left(\frac{\hat{p}_x^2}{2m}-\mu\lambda_z^{\bm{J}}\right)\tau_z
-\widehat{\bm{M}}(x)\cdot\left(\tau_z\sigma_x,\sigma_y,\tau_z\sigma_z\right)-\Delta\tau_y\sigma_y\,,
\label{eq:Gauged Hamiltonian magnetism confined}
\eea

where $\widehat{\bm{M}}(x)$ is a real matrix defined in $\left| n\right>$ space. If $\bm{J}=\bm{0}$ and $\widehat{M}_y(x)=0$, ${\cal K}$ is a symmetry of the Hamiltonian
and the system belongs to class BDI \cite{comment}. Instead, if $\widehat{M}_y(x)\neq0$, the system belongs to class D. For the special case where $\bm{M}(\bm{r})$ does not
depend on the $y$-coordinate, i.e. $\bm{M}(\bm{r})=\bm{M}(x)$, $\widehat{\bm{M}}(\bm{r})$ becomes diagonal and can be divided into an infinite number of sub-spaces labelled by
$n$ yielding
\bea
\widehat{{\cal H}}_n'(\hat{p}_x{\color{black},x})=\left[\frac{\hat{p}_x^2}{2m}-\mu+\hbar\omega\left(n+\frac{1}{2}\right)\right]\tau_z
-\bm{M}(x)\cdot\left(\tau_z\sigma_x,\sigma_y,\tau_z\sigma_z\right)-\Delta\tau_y\sigma_y\,,
\eea

which leads to the total symmetry class $\oplus_n$BDI. If $\bm{M}(x)$ violates ${\cal K}$ we obtain a direct sum $\oplus_n$D. By allowing a finite $\bm{J}$ we also violate
${\cal K}$. Specifically, if $\bm{J}=\left(J_x,0\right)$, that corresponds to the case $\Delta(\bm{r})=\Delta(x)$, the system resides in the class $\oplus_n$D. However, if
$\bm{J}=\left(0,J_y\right)$ the system belongs to class D \cite{comment}, due to the simultaneous presence of $\lambda_y$ and $\lambda_z^{\bm{J}}$ in the Hamiltonian of
Eq.~(\ref{eq:Gauged Hamiltonian magnetism confined}), that do not allow the decomposition in $n$-sectors. In Table \ref{table:Topological Phases} the general case where
$\bm{M}$ and $\Delta$ depend on both coordinates is presented.

\noi \bt{1d system:} By dimensional reduction on the Hamiltonian of Eq.~(\ref{eq:Gauged Hamiltonian magnetism}) we obtain the following pure 1d model
\bea
\widehat{{\cal H}}'(\hat{p}_x{\color{black},x})=\frac{\hbar}{2m}\ph J_x\hat{p}_x I+\left[\frac{\hat{p}_x^2}{2m}+
\frac{\left(\hbar J_x/2\right)^2}{2m}-\mu\right]\tau_z-\bm{M}(x)\cdot\left(\tau_z\sigma_x,\sigma_y,\tau_z\sigma_z\right)-\Delta\tau_y\sigma_y\,.
\label{eq:1d model for magnetism and superconductivity}
\eea

If $M_y(x)=0$ and $J_x=0$, ${\cal K}$ is conserved and the system belongs to class BDI. Instead, if one of the previous terms is non-zero, the Hamiltonian is not real any
more and it falls into symmetry class D \cite{Choy}.\\

\item \bt{Cases \RNum{5} and \RNum{6}}

In the last part of this section I complete the possible cases by considering the situation where all the terms of Eq.~(\ref{eq:Hamiltonian}) are present. The latter
equation in combined Nambu and spin spaces reads
\bea
\widehat{{\cal H}}(\hat{\bm{p}},\bm{r})&=&
\left[\frac{\hat{\bm{p}}^2}{2m}-\mu+V(\bm{r})\right]\tau_z+\frac{\left\{v(\bm{r}),\hat{p}_x\tau_z\sigma_y-\hat{p}_y\sigma_x\right\}}{2}
-\bm{M}(\bm{r})\cdot{\color{black}\left(\tau_z\sigma_x,\sigma_y,\tau_z\sigma_z\right)}-\Delta_{\Re}(\bm{r})\tau_y\sigma_y\no\\&-&\Delta_{\Im}(\bm{r})\tau_x\sigma_y\,.
\eea

When magnetism and Rashba spin-orbit coupling coexist, the accessible topological phases constitute an overlap of the previously examined separate cases. Therefore here
we will investigate what are the consequences of the addition of magnetism in Cases \RNum{1} and \RNum{2} for different dimensionalities. Earlier, we observed that when
magnetism is not present, there are two possible scenarios depending on the behaviour of the superconducting order parameter under ${\cal K}$. 

\noi \bt{2d and quasi-1d systems:} For the specific cases, if $\Delta(\bm{r})=\Delta_{\Re}(\bm{r})$ the system resides in the symmetry class DIII being invariant under ${\cal
T}$. If $\bm{M}(\bm{r})$ is introduced, ${\cal T}$ will be broken and the system will transit to class D. If $\Delta(\bm{r})$ is complex, the system is already in class D, and
consequently the inclusion of magnetism leads to no additional effects. 

\noi \bt{1d system:} For pure 1d systems the presence of a magnetic order is crucial and leads to new TSC phases. The 1d descendant of the above Hamiltonian reads
\bea
\widehat{{\cal H}}(\hat{p}_x,x)=
\left(\frac{\hat{p}_x^2}{2m}-\mu\right)\tau_z+\frac{\left\{v(x),\hat{p}_x\right\}}{2}\ph\tau_z\sigma_y
-\bm{M}(x)\cdot{\color{black}\left(\tau_z\sigma_x,\sigma_y,\tau_z\sigma_z\right)}-\Delta_{\Re}(x)\tau_y\sigma_y-\Delta_{\Im}(x)\tau_x\sigma_y\,.
\eea

From Table~\ref{table:Topological Phases} we immediately observe that {\color{black}no MFs emerge fundamentally} in the absence of magnetism. As mentioned earlier, the reason is
that the presence of the spin-orbit coupling term $\tau_z\sigma_y$ alone, cannot lock the spin-quantization axis. Nevertheless, the addition of a perpendicular magnetization
field remedies this problem and can lead to TSC phases {\color{black}with MFs}. If {\color{black}$\Delta(x)$ and $\bm{M}(x)$ are invariant under ${\cal K}$}, the Hamiltonian is
characterized by a generalized time-reversal symmetry $\Theta={\cal K}$ and a chiral symmetry $\Pi=\tau_x$ which permits an integer number of MFs per edge \cite{Tewari and
Sau}. The translationally invariant version of this model
\bea
\widehat{{\cal H}}(\hat{p}_x)=
\left(\frac{\hat{p}_x^2}{2m}-\mu\right)\tau_z+v\hat{p}_x\tau_z\sigma_y-\bm{M}\cdot{\color{black}\left(\tau_z\sigma_x,\sigma_y,\tau_z\sigma_z\right)}-\Delta\tau_y\sigma_y\,,
\label{eq:MF-wire}
\eea

corresponds to the celebrated MF-wire proposal \cite{Oreg} which currently under intense experimental investigation \cite{MF experiments} and concerns a Rashba
semiconducting wire in the presence of a Zeeman field and proximity induced superconductivity. The systems transits to the topologically non-trivial phase when the criterion 
\bea |\bm{M}|>\sqrt{\mu^2+\Delta^2}\label{eq:criterion}\,,
\eea

is satisfied. Finally, if {\color{black}$\Delta_{\Im}(x)$ or (and)} $M_y(x)\neq0$ then ${\cal K}$ is broken and the system belongs to class D with a $\mathbb{Z}_2$ invariant
allowing for a single MF per edge.

\end{itemize}

\section{Topological superconductivity based on spin-orbit coupling and supercurrents in the absence of magnetism}

In Case \RNum{2}, I showed that a {\color{black}quasi-1d} system characterized by Rashba spin-orbit coupling and ${\cal T}$-breaking superconductivity belongs to symmetry class
D, which can in principle support MFs, without any kind of magnetism. In this paragraph I explicitly demonstrate that this scenario is feasible and experimentally accessible.
Here I will consider a heterostructure consisting of conventional superconductors in proximity to \bt{A.} the surface of a 3d topological insulator (TI) and \bt{B.} a
double-Rashba semiconducting wire setup, which constitutes the simplest example of a {\color{black}quasi-1d semiconductor}. In both cases, the additional presence of a finite
supercurrent, will be crucial for engineering topological superconductivity.

{\color{black}
\subsection{Topological superconductor (TSC) in a heterostructure consisting of a TI and conventional SCs}
}
The respective Hamiltonian describing the TI surface states in the presence of induced pairing reads 
\bea
\widehat{{\cal H}}(\hat{\bm{p}},\bm{r})=
-\mu\tau_z+v\left(\hat{p}_x\tau_z\sigma_y-\hat{p}_y\sigma_x\right)-\Delta_{\Re}(\bm{r})\tau_y\sigma_y-\Delta_{\Im}(\bm{r})\tau_x\sigma_y\,,
\eea

\noi which is derived from Eq.~(\ref{eq:Hamiltonian Rashba and superconductivity}) by considering $v(\bm{r})=v$, $V(\bm{r})=0$ and $m\rightarrow+\infty$. In fact, the latter
model can be linked to a previous proposal Ref.~\cite{Fu and Kane}. Notice that I permitted a particle-hole asymmetric bulk TI by allowing a finite chemical potential, which
additionally ensures that the system resides in class D. Nevertheless, for the rest of the discussion, I will for simplicity set $\mu=0$. The latter special case, enhances the
symmetry of the system leading to the following symmetry class transition ${\rm D}\rightarrow{\rm BDI}$, due to the emergence of a chiral symmetry with matrix $\sigma_z$,
without though affecting our analysis concerning the emergence of MFs. At this point I include a finite supercurrent along the $y$-axis by considering
$\Delta(\bm{r})=\Delta(y)=\Delta e^{iJy}$. Furthermore, I assume that $Jy$ is small which allows us to make the approximation $\Delta(y)\simeq\Delta+i\Delta Jy$. Under these
assumptions the Hamiltonian becomes
\bea
\widehat{{\cal H}}(\hat{\bm{p}},y)=
v\left(\hat{p}_x\tau_z\sigma_y-\hat{p}_y\sigma_x\right)-\Delta\tau_y\sigma_y-\Delta Jy\tau_x\sigma_y\,.\label{eq:TI}
\eea

\begin{figure}[b]
\includegraphics[scale=0.17]{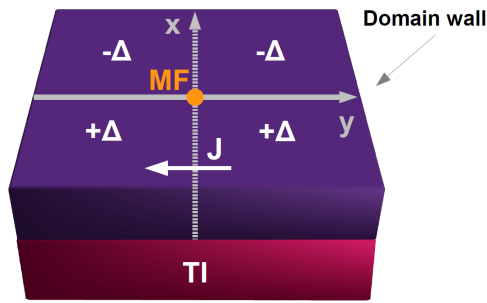}
\caption{A heretostructure consisting of conventional superconductors deposited on top of the surface of a 3d topological insulator. The presence of a superconducting
Josephson junction of supercurrent $\bm{J}=(0,J)$, in combination with a $\pi$-superconducting phase domain wall, along $x=0$, traps a MF at $\bm{r}=(0,0)$.}
\label{fig:TISC}
\end{figure}

By squaring the BdG Hamiltonian operator, we obtain
\bea
\widehat{{\cal H}}^2(\hat{\bm{p}},y)=(v\hat{\bm{p}})^2+\Delta^2+(\Delta J)^2y^2+v\hbar\Delta J\tau_x\sigma_z\,.
\eea

The above Hamiltonian can be diagonalized in the $y,\hat{p}_y$ space by introducing the eigen-states $\left|n\right>$ of a quantum harmonic oscillator with frequency
$\omega=2v\Delta J$ providing 
\bea
\widehat{{\cal H}}_n^2(\hat{p}_x)=(v\hat{p}_x)^2+\Delta^2+v\hbar\Delta J(2n+1)+v\hbar\Delta J\tau_x\sigma_z\,.
\eea

Notice that the presence of the supercurrent leads to confinement parallel to its direction. Since we are interested in the low energy regime, we can restrict to the
eigen-states of $\tau_x\sigma_z$ with eigen-value $-1$ and $n=0$. In fact, for the latter eigenstates, the term $v\hbar\Delta J(2n+1)+v\hbar\Delta J\tau_x\sigma_z$
becomes zero, rendering these solutions as Majorana bound state solutions in the absence of $(v\hat{p}_x)^2+\Delta^2$. With this in mind, I project the following part
${\color{black}v\hat{p}_x\tau_z\sigma_y-\Delta\tau_y\sigma_y}$ of Eq.~(\ref{eq:TI}) onto these degenerate lowest energy states, leading to the effective Hamiltonian
\bea
\widehat{{\cal H}}_{eff}(\hat{p}_x)=v\hat{p}_x\kappa_y+\Delta\kappa_x\,,
\eea

where $\bm{\kappa}$ correspond to Pauli matrices defined in the truncated basis spanned by the Majorana bound states \mbox{$\left|n=0;\tau_x=-1;\sigma_z=+1\right>$} and
\mbox{$\left|n=0;\tau_x=+1;\sigma_z=-1\right>$.} The latter results are in absolute agreement with the SC-TI-SC heterostructure model considered in Ref.~\cite{Fu and Kane} and
related studies concerning graphene-based hybrid devices \cite{Graphene}, following a different approach. {\color{black} Fu and Kane \cite{Fu and Kane}}, considered a
tri-junction of SC-TI-SC systems in order to implement a $C_3$ vortex at the meeting point which can host a MF. In fact, the SC-TI-SC setup has been recently under
experimental investigation \cite{Analytis} revealing possible signatures of MFs. Here, for the detection of MFs, I propose the situation of a $\pi$-phase domain wall for the
SC gap $\Delta$ along the $x$-axis (Fig.~\ref{fig:TISC}), in analogy to the Jackiw-Rebbi model \cite{Jackiw}. However, in the present case the bound states will be of the
Majorana type. Note that the  equivalent description of the SC-TI-SC heterostructure proposed in Ref.~\cite{Fu and Kane}, using supercurrents as in the present discussion, had
not been so far realized, leaving alternative accessible MF setups unexplored. According to the analysis above, a prominent system for hosting MFs is a quasi-1d Rashba
semiconductor in proximity to a conventional superconductor. As I demonstrate in the next paragraph, the presence of a Josephson current flow parallel to the direction where
confinement is imposed, will lead to the appearance of edge MFs.

{\color{black}
\subsection{TSC in a heterostructure consisting of two coupled Rashba semiconducting wires and conventional SCs}
}
In this subsection I will focus on quasi-1d Rashba semiconducting platforms. Due to the quasi-1d character of the system, a finite number of channels {\color{black}is} generally
allowed, which should be taken into full consideration for the MF analysis. Nevertheless, in order to demonstrate the possibility of MFs, based solely on supercurrents, I will
here consider the simplest example of a quasi-1d Rashba semiconductor, which consists of two coupled single channel wires (Fig.~\ref{fig:Doublewire}). Note that double-wire
setups \cite{doubleWire} have been recently considered in the context of ${\cal T}$-invariant TSCs. However, in our case ${\cal T}$ will be broken. The relevant Hamiltonian
reads
\bea
{\cal H}&=&\int
dx\sum_{n=\pm}\left\{\hat{\psi}_n^{\dag}(x)\left(\frac{\hat{p}_x^2}{2m}-\mu+v\hat{p}_x\sigma_y\right)\hat{\psi}_n\up(x)+
\left[\Delta e^{ni\frac{\delta\varphi}{2}}\psi_{\uparrow,n}^{\dag}(x)\psi_{\downarrow,n}^{\dag}(x)+{\rm h.c.}\right]\right\}\no\\
&+&\int dx\left\{t_{\perp}\up\hat{\psi}_+^{\dag}(x)\hat{\psi}_-\up(x)+iV_{\perp}\up\hat{\psi}_+^{\dag}(x)\sigma_x\hat{\psi}_-\up(x)+
\Delta_{\perp}\up\left[\psi_{\uparrow,+}^{\dag}(x)\psi_{\downarrow,-}^{\dag}(x)+\psi_{\uparrow,-}^{\dag}(x)\psi_{\downarrow,+}^{\dag}(x)\right]+{\rm h.c.}\right\}\,,
\label{eq:2wiresHamiltonian}
\eea

\begin{figure}[t]
\includegraphics[scale=0.3]{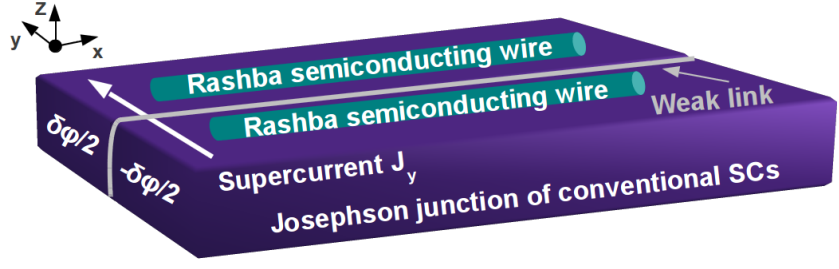}
\caption{A heretostructure consisting of two coupled single channel wires deposited on top of two conventional superconductors interfaced by a weak link permitting the flow
of a Josephson current. The supercurrent flow is directed transversely to the wires' axis and is sufficient to generate edge MFs that are ``shared'' by the two wires.}
\label{fig:Doublewire}
\end{figure}

where $n=\pm$ labels the two parallel single-channel wires placed at distance $L_y$, while $t_{\perp}$, $V_{\perp}$, $\Delta$ and $\Delta_{\perp}$ correspond to inter-wire
hopping, inter-wire spin-orbit coupling, intra-wire superconductivity and inter-wire superconductivity respectively. Moreover, I also introduced a finite supercurrent
$J_y\sim\delta\varphi$ flowing from one wire to the other, by incorporating a phase in the intra-wire superconducting gap that has an opposite sign on the two wires. Notice
that the inter-{\color{black}wire} superconducting term is unaffected by the presence of the supercurrent for the particular direction of flow. For a compact description, I will
introduce the spinor 
\bea
\widehat{\Psi}^{\dag}(x)=
\left(\psi_{\uparrow,+}^{\dag}(x)\,,\psi_{\downarrow,+}^{\dag}(x)\,,\psi_{\uparrow,-}^{\dag}(x)\,,\psi_{\downarrow,-}^{\dag}(x)\,,
\psi_{\uparrow,+}\up(x)\,,\psi_{\downarrow,+}\up(x)\,,\psi_{\uparrow,-}\up(x)\,,\psi_{\downarrow,-}\up(x)\right)\,,
\eea

and additionaly employ the $\bm{\kappa}$ Pauli matrices that act on the subspace spanned by the two wire indices $n=\pm$. The BdG Hamiltonian of
Eq.~(\ref{eq:2wiresHamiltonian}) reads
\bea
\widehat{{\cal H}}(\hat{p}_x)=\left(\frac{\hat{p}_x^2}{2m}-\mu\right)\tau_z+t_{\perp}\tau_z\kappa_x+v\hat{p}_x\tau_z\sigma_y-V_{\perp}\up\kappa_y\sigma_x
-\Delta e^{i\frac{\delta\varphi}{2}\tau_z\kappa_z}\tau_y\sigma_y-\Delta_{\perp}\up\tau_y\kappa_x\sigma_y\,.
\label{eq:2wiresHamiltonianSpinor}
\eea

It is straighforward to confirm that the above Hamiltonian is characterized by a chiral symmetry with matrix $\tau_x\kappa_x$ and a concomitant generalized time-reversal
symmetry $\Theta=\kappa_x{\cal K}$. Due to the property $\Theta^2={\rm I}$, the system resides in class {\rm BDI} which in 1d is characterized by a $\mathbb{Z}$ topological
invariant, allowing an integer number of topologically protected MFs per edge \cite{Tewari and Sau,Lutchyn and Fisher}. For the rest I will consider $\mu=0$ which can be
always experimentally achieved by properly gating the device and does not affect our analysis. It is instructive to study the energy spectrum for $p_x=0$ when the supercurrent
is zero, which reads
\bea
E_{\delta\varphi=0}(p_x=0)=\pm\sqrt{t_{\perp}^2+\left(\Delta_{\perp}\up\pm\sqrt{V_{\perp}^2+\Delta^2}\right)^2}\,.
\eea

We observe that the spectrum is twofold degenerate and the only possibility for a gap closing at $p_x=0$, which would imply the presence of MFs, can occur only if
$t_{\perp}=0$ and $\Delta_{\perp}\up=\sqrt{V_{\perp}^2+\Delta^2}$. However, even if we consider $t_{\perp}=0$, for every realistic case $\Delta>\Delta_{\perp}$. Consequently,
in the absence of a supercurrent, the system cannot support MFs. In order to shed light on how the presence of a finite supercurrent can lead to MFs, I will perform a gauge
transformation, $\widehat{{\cal H}}'(\hat{p}_x)\equiv e^{-i\frac{\delta\varphi}{4}\tau_z\kappa_z}\widehat{{\cal H}}(\hat{p}_x)e^{i\frac{\delta\varphi}{4}\tau_z\kappa_z}$, in
order to remove the superconducting phase and obtain an expression similar to Eq.~(\ref{eq:Gauged Hamiltonian}). Furthermore, I will consider $\delta\varphi=\pi-2\epsilon$
where $\epsilon$ is considered small and I will keep terms linear in $\epsilon$. Under these conditions, the Hamiltonian of Eq.~(\ref{eq:2wiresHamiltonianSpinor}) becomes
\bea
\widehat{{\cal H}}'(\hat{p}_x)=\frac{\hat{p}_x^2}{2m}\tau_z+t_{\perp}\kappa_y+\epsilon t_{\perp}\tau_z\kappa_x+v\hat{p}_x\tau_z\sigma_y
+V_{\perp}\tau_z\kappa_x\sigma_x-\epsilon V_{\perp}\up\kappa_y\sigma_x-\Delta\tau_y\sigma_y-\Delta_{\perp} \up\tau_y\kappa_x\sigma_y\,.
\label{eq:2wiresHamiltonianGauged}
\eea

Notice that in the presence of a supercurrent for which $\epsilon=0$, the inter-wire spin-orbit coupling term $V_{\perp}\up\kappa_y\sigma_x$ is converted completely
into an inter-wire Zeeman term $V_{\perp}\tau_z\kappa_x\sigma_x$, which is polarized perpendicular to the intra-wire spin-orbit coupling term $v\hat{p}_x\tau_z\sigma_y$ and is
crucial for the appearance of MFs in this double-wire setup. For $\epsilon=0$, the reconstructed energy spectrum for $p_x=0$ reads
\bea
E_{\delta\varphi=\pi}(p_x=0)=\pm\left[\Delta\pm\sqrt{t_{\perp}^2+\left(V_{\perp}\pm \Delta_{\perp}\right)^2}\right]\,.\label{eq:spectrumm}
\eea

We observe that there is no-degeneracy at $p_x=0$, which implies that we obtain a single MF per gap closing. For the above spectrum there can be two gap closings at $p_x=0$
occuring for $\Delta=\sqrt{t_{\perp}^2+\left(V_{\perp}\pm \Delta_{\perp}\right)^2}$ marking the related topological phase boundaries. According to the latter analysis and by
additionally calculating the related $\mathbb{Z}$ topological invariant, following Ref.~\cite{Tewari and Sau}, I find that the system resides in the topologically non-trivial
phase with a single edge MF when the criterion $\sqrt{t_{\perp}^2+\left(V_{\perp}-\Delta_{\perp}\right)^2}<\Delta<\sqrt{t_{\perp}^2+\left(V_{\perp}+\Delta_{\perp}\right)^2}$
is satisfied. 

To illustrate the appearance of MFs in a transparent way, I will consider first the following special case $\epsilon=t_{\perp}=0$, where the Hamiltonian of
Eq.~(\ref{eq:2wiresHamiltonianGauged}) enjoys a unitary symmetry generated by the matrix $\kappa_x$ which implies that the two wires are mirror symmetric. The particular
mirror symmetry allows for the block diagonalization of the Hamiltonian in the following manner
\bea
\widehat{{\cal H}}_{\kappa}'(\hat{p}_x)=\frac{\hat{p}_x^2}{2m}\tau_z+v\hat{p}_x\tau_z\sigma_y+\kappa V_{\perp}\tau_z\sigma_x-(\Delta+\kappa\Delta_{\perp}\up)\tau_y\sigma_y\,,
\label{eq:2wiresHamiltonianBlock}
\eea

where $\kappa=\pm1$ correspond to the eigen-values of $\kappa_x$. The Hamiltonian of each block is essentially the Hamiltonian of the strictly 1d wire TSC discussed in
Ref.~\cite{Sau,Oreg}, which belongs to class {\rm BDI}, and supports a single MF per edge when the following criterion is satisfied $V_{\perp}>\Delta+\kappa\Delta_{\perp}$. 
Consequently, the system resides: \bt{a.} in the topologically trivial phase for $V_{\perp}<\Delta-\Delta_{\perp}$, \bt{b.} in the topologically non-trivial phase with a
single MF for $|V_{\perp}-\Delta|<\Delta_{\perp}$ and \bt{c.} in the topologically non-trivial phase with two MFs for $V_{\perp}>\Delta+\Delta_{\perp}$. It is desirable to
study the fate of the MFs when the additional chiral symmetry (with matrix $\tau_x$) is broken and a symmetry class transition {\rm BDI}$\oplus${\rm BDI}$\rightarrow${\rm BDI}
occurs for the Hamiltonian of Eq.~(\ref{eq:2wiresHamiltonianBlock}), due to a finite $\epsilon$. For this purpose, I construct the following low energy effective model 
\bea
\widehat{{\cal H}}_{eff}'(\hat{p}_x)=v\hat{p}_x\rho_y+(V_{\perp}-\Delta)\eta_z\rho_x-\Delta_{\perp}\rho_x\,,\label{eq:effectiveI}
\eea
by projecting the Hamiltonian of Eq.~(\ref{eq:2wiresHamiltonianBlock}) onto the following gap closing related Majorana bound state solutions:
\bea
\left|1\right>=\left|\tau_x=+1;\kappa_x=+1;\sigma_z=+1\right>\,,&&\left|2\right>=\left|\tau_x=-1;\kappa_x=+1;\sigma_z=-1\right>\,,
\no\\
\left|3\right>=\left|\tau_x=-1;\kappa_x=-1;\sigma_z=+1\right>\,,&&\left|4\right>=\left|\tau_x=+1;\kappa_x=-1;\sigma_z=-1\right>\,.\eea

For the latter procedure I neglected the quadratic in momentum kinetic term $\sim\hat{p}_x^2\tau_z$ since I focus on momenta about $p_x=0$, while I made use of the
$\bm{\eta}$ (acting on $\kappa_x=\pm1$ blocks) and $\bm{\rho}$ Pauli matrices. The spectrum of the effective model has the following form
\bea
E_{\delta\varphi=\pi}(p_x)=\pm\sqrt{(vp_x)^2+\left(V_{\perp}-\Delta\pm\Delta_{\perp}\right)^2}\,,
\eea

owing the anticipated gap closings at $V_{\perp}=\Delta\pm\Delta_{\perp}$, which provide the topological phase boundaries. At this point, I assume that $\epsilon$ is small.
By adding the corresponding term $-\epsilon V_{\perp}\kappa_y\sigma_x$ as a perturbation to the above effective model, I finally obtain
\bea
\widehat{{\cal H}}_{eff}'(\hat{p}_x)=v\hat{p}_x\rho_y+(V_{\perp}-\Delta)\eta_z\rho_x-\Delta_{\perp}\rho_x+\epsilon V_{\perp}\eta_y\rho_x\,.\label{eq:effective}
\eea

The inclusion of $\epsilon$ modifies crucially the energy spectrum, which now reads 
\bea
E_{\delta\varphi\neq\pi}(p_x)=\pm\sqrt{(vp_x)^2+\left[\sqrt{(V_{\perp}-\Delta)^2+(\epsilon V_{\perp})^2}\pm\Delta_{\perp}\right]^2}\,.
\eea

We directly observe that there is only one possible gap closing and consequently only one accessible topological phase supporting a single MF per edge. This can be naturally
understood by taking into consideration that the two MFs, previously existing for the topologically non-trivial phase with $V_{\perp}>\Delta+\Delta_{\perp}$, hybridize and
give rise to a finite energy fermionic solution. Note that the chiral symmetry breaking effects are non-perturbative. In fact, we may observe the effect of an infinitessimal
$\epsilon$ by rewriting the energy spectrum in the following form 
\bea
E_{\delta\varphi\neq\pi}(p_x)=\pm\sqrt{(vp_x)^2+\left(|V_{\perp}-\Delta|\pm\Delta_{\perp}\right)^2}\,.
\eea

We notice that an infinitessimal $\epsilon$ will merge the previous three distinct phases of zero, one or two MFs into the following two: \bt{a.} a topologically trivial
superconducting phase for $|V_{\perp}-\Delta|>\Delta_{\perp}$ and \bt{b.} a topologically non-trivial phase with a single MF per edge for $|V_{\perp}-\Delta|<\Delta_{\perp}$.

In order to make a connection to the related experimental setup, I will consider {\rm InSb} wires for which we have $v\hbar=0.2{\rm eV\AA}$, $m=0.015m_e$ and $\Delta=250{\rm
\mu eV}$. Furthermore, $t_{\perp}\simeq \hbar^2/(2mL_y^2)$ and $V_{\perp}\simeq v\hbar/L_y$. By assuming a constant value for $\Delta_{\perp}\sim \Delta/5=50{\rm \mu eV}$,
Eq.~(\ref{eq:spectrumm}) and also the computation of the related topological invariant provide that the system resides in the topologically non-trivial phase with
a single MF per edge for $109{\rm nm}<L_y<131{\rm nm}$. In this regime we expect a zero-bias anomaly peak in the tunneling spectra, which could constitute a sharp signature of
MF physics.\\

\section{Examples of topological phases with hidden symmetries}

\noi In this section I will {\color{black}present} two examples where unitary or anti-unitary hidden symmetries occur for some of the TSC phases presented in Table
\ref{table:Topological Phases} and demonstrate what are the concomitant modifications of the initial symmetry class.

\begin{itemize}

\item \bt{Cases \RNum{1} and \RNum{2} in the presence of a single unitary hidden symmetry} $\bm{{\cal O}}_u$

\noi Let us now investigate the consequences of the emergence of a ``hidden'' symmetry due to the special form of the Rashba spin-orbit coupling term. As a case study I will
focus on the topological properties of the following quasi-1d {\color{black}Hamiltonian}, introduced in Eq.~(\ref{eq:Hamiltonian quasi-1d random Rashba}) of
Case \RNum{1}
\bea
\widehat{{\cal H}}(\hat{p}_x{\color{black},x})&=&\frac{\hbar}{2m}\left(J_x\hat{p}_xI+J_y\hbar\lambda_y\right)+\left(\frac{\hat{p}^2_x}{2m}-\mu\lambda_z^{\bm{J}}\right)\tau_z+
\frac{\{v(x),\hat{p}_x\}}{2}\ph\tau_z\sigma_y-v(x)\hbar\lambda_y\sigma_x+\frac{v(x)\hbar}{2}\left(J_x\sigma_y-J_y\tau_z\sigma_x\right)\no\\&-&\Delta\tau_y\sigma_y\,.
\eea

Here we will restrict to the special situation where $t_{\pi/Q}v(x)=v(x+\pi/Q)=-v(x)$. We may readily observe in which manner this property leads to an emergent unitary
symmetry. The terms of the Hamiltonian that do not contain $v(x)$ are invariant under arbitrary translations and under the action of $\sigma_h$, which in our formalism is
represented as $\sigma_h=i\tau_z\sigma_z$ in spin-space. Since all the terms with coefficient $v(x)$ are odd under $\sigma_h$, the full Hamiltonian is invariant under the
action of ${\cal O}_u=\sigma_ht_{\pi/Q}$. The appearance of a hidden symmetry ${\cal O}_u$ leads to an additional generalized time-reversal symmetry $\widetilde{\Theta}={\cal
O}_u{\cal T}$ and a concomitant chiral symmetry $\widetilde{\Pi}={\cal O}_u\Pi${\color{black}, when $\bm{J}=\bm{0}$}. The emergence of ${\cal O}_u$ modifies the symmetry class
of the system by splitting the symmetry classes DIII and D found earlier, into DIII$\oplus$DIII and D$\oplus$D, respectively. {\color{black}Note that point group symmetry
protected phases are currently under intense investigation \cite{Point Group} and a topological classification of systems with reflection symmetry has also appeared
\cite{Shinsei}}. A simple example for $v(x+\pi/Q)=-v(x)$ is $v(x)=2v_Q\up\cos(Qx+\theta)$ with $Q=2q$. Here, $\theta$ is considered pinned to a constant value. The modulated
spin-orbit coupling term can be viewed as an unconventional spin triplet density wave \cite{UDW}, similar to the Rashba spin-orbit density wave proposed in \cite{Das Tanmoy}
as a potential candidate for the so called ``hidden order'', which appears in the non-superconducting regime of the phase diagram of the heavy fermion compound ${\rm
URu_2Si_2}$.
 
The simultaneous presence of the momentum operator $\hat{p}_x$ and coordinate $x$ does not allow for a direct and transparent inspection of further topological
properties of the system. Nonetheless, it is also possible in principle to obtain through some kind of ``deformation'' procedure (in the topological sense) a model defined
solely in momentum space that shares the same symmetries and topological properties with the original model. In order for this mapping to be meaningful and offer a direct
computation of topological invariants, translational symmetry must be somehow restored. The presence of a periodic $v(x)=2v_Q\up\cos(Qx+\theta)$ term, leads to the formation
of a band structure with a Brillouin zone of length $Q$ since $t_{2\pi/Q}v(x)=v(x)$. The property $t_{\pi/Q}v(x)=-v(x)$ gives rise to a sublattice structure that will
eventually lead to the two sub-blocks of the Hamiltonian that become relevant in the presence of ${\cal O}_u$. Since we are not interested in the full band structure, but
mainly {\color{black}in removing the $x$-dependence of the Hamiltonian}, we may expand the field operator in the following fashion
\bea
\hat{\psi}(x)\simeq e^{+iqx}\hat{\psi}_{+q}(x)+e^{-iqx}\hat{\psi}_{-q}(x)\,,\label{eq:spinor expansion}
\eea

\noi where $\hat{\psi}_{\pm q}(x)$ are slowly varying fields leading to the Hamiltonian
\bea
\widehat{{\cal H}}_{q}(\hat{p}_x)&=&\frac{\hbar}{2m}\left(J_x\hat{p}_xI+J_y\hbar\lambda_y\right)+\frac{\hbar}{2}\tilde{v}J_x\rho_z
+\left[\frac{\hat{p}^2_x}{2m}+\frac{(\hbar q)^2}{2m}-\mu\lambda_z^{\bm{J}}\right]\tau_z+\tilde{v}\hat{p}_x\tau_z\rho_z\no\\
&&+
v_Q\up\hat{p}_x\left(\cos\theta\tau_z\rho_x\sigma_y-\sin\theta\tau_z\rho_y\sigma_y\right)-
v_Q\up\hbar\left(\cos\theta\rho_x\lambda_y\sigma_x-\sin\theta\rho_y\lambda_y\sigma_x\right)\no\\
&&+
\frac{v_Q\up\hbar}{2}J_x\left(\cos\theta\rho_x\sigma_y-\sin\theta\rho_y\sigma_y\right)-
\frac{v_Q\up\hbar}{2}J_y\left(\cos\theta\tau_z\rho_x\sigma_x-\sin\theta\tau_z\rho_y\sigma_x\right)-\Delta\tau_y\sigma_y\,,\label{eq:qq}
\eea

with $\tilde{v}=\hbar q/m$. The above Hamiltonian acts on the enlarged spinor
\bea
\widehat{\Psi}^{\dag}_{q}(x)=\left(
\psi_{+q\uparrow}^{\dag}(x)\,,\psi_{+q\downarrow}^{\dag}(x)\,,\psi_{-q\uparrow}^{\dag}(x)\,,\psi_{-q\downarrow}^{\dag}(x)\,,
\psi_{-q\uparrow}\up(x)\,,\psi_{-q\downarrow}\up(x)\,,\psi_{+q\uparrow}\up(x)\,,\psi_{+q\downarrow}\up(x)\right)\,,
\eea

with the $\bm{\rho}$ Pauli matrices acting on $\pm q$ space. Notice that terms with $\rho_x$ or $\rho_y$ carry momentum $Q$. By expanding the field operator in
this manner, we managed to end up with a {\color{black}coordinate independent} Hamiltonian. This approximation allows us to readily study topological aspects in momentum space
which for the specific case is an easier task compared to the required analysis in coordinate space. Nonetheless, it is not a priori ensured that the coordinate and momentum
pictures are topologically equivalent. If they do, this approximation constitutes a suitable deformation procedure for mapping $x$ to $k_x$ space topology.

In order to confirm if these systems belong to the same symmmetry class, we have to study the emerging symmetries for the latter model. In this basis the expression for
the generalized time-reversal symmetry operator simplifies to $\Theta=i\sigma_y{\cal K}=i\rho_x\sigma_y{\cal K}'$, where ${\cal K}'$ is a complex conjugation operator not
acting on $Q$ or $q$. The presence of $\rho_x$ effects complex conjugation operation ${\cal K}$ in $q$-space, since $q\rightarrow-q$ is equivalent to ${\cal K}=\rho_x{\cal
K}'$. Furthermore, within the specific framework {\color{black}$t_{\pi/Q}=-i\rho_z$}. Notice that $(-i\rho_z)^2=-{\rm I}$, i.e. similar to the behaviour of rotation operators
for a spin-1/2. This is a direct consequence of the fact that the spinor contains the wave-vector $q$ which is half of the wave-vector $Q=2q$. We may directly confirm that
$[\widehat{{\cal H}}_{q}(\hat{p}_x),\sigma_ht_{\pi/Q}]\equiv[\widehat{{\cal H}}_{q}(\hat{p}_x),\tau_z\rho_z\sigma_z]=0$. When $\bm{J}=\bm{0}$, the Hamiltonian is invariant
under $\Theta$ and the presence of ${\cal O}_u=\sigma_ht_{\pi/Q}$ leads to the additional time-reversal symmetry $\widetilde{\Theta}={\cal
O}_u\Theta=i\tau_z\rho_y\sigma_x{\cal K}'$. The emerging chiral symmetries for this model read $\Pi=\tau_x\sigma_y$ and $\widetilde{\Pi}=\tau_y\rho_z\sigma_x$. As expected, in
this case the system belongs to the symmetry class DIII$\oplus$DIII. Furthermore, if $\bm{J}\neq\bm{0}$ then $\Theta$ is broken and the system transits to the symmetry class
D$\oplus$D. To explicitly demonstrate the sub-block structure of the Hamiltonian and the direct sum of symmetry classes, I effect the unitary transformation
\bea
\Lambda=\frac{{\rm I}+\rho_z\sigma_z}{2}+\tau_x\frac{{\rm I}-\rho_z\sigma_z}{2}\,,
\eea

which transforms the Hamiltonian as follows $\widehat{{\cal H}}_{q}^{rot}(\hat{p}_x)=\Lambda^{\dag}\widehat{{\cal H}}_{q}(\hat{p}_x)\Lambda$. This particular
unitary transformation block diagonalizes the matrix $\tau_z\rho_z\sigma_z$, representing the hidden symmetry operation ${\cal O}_u$, into
$\Lambda^{\dag}\tau_z\rho_z\sigma_z\Lambda=\tau_z$. The transformed Hamiltonian is block diagonal and can be labelled by the eigenvalues of $\tau_z$, $\tau=\pm 1$,
yielding
\bea
\widehat{{\cal H}}_{q,\tau}^{rot}(\hat{p}_x)&=&
\frac{\hbar}{2m}\left(J_x\hat{p}_x+J_y\hbar\lambda_y\right)I+\frac{\hbar}{2}\tilde{v}J_x\rho_z
+\tau\left[\frac{\hat{p}^2_x}{2m}+\frac{(\hbar q)^2}{2m}-\mu\lambda_z^{\bm{J}}\right]\rho_z\sigma_z+\tau\tilde{v}\hat{p}_x\sigma_z\no\\
&&
+v_Q\up\left(\tau\hat{p}_x\cos\theta+\hbar\lambda_y\sin\theta\right)\rho_y\sigma_x+v_Q\up\left(\tau\hat{p}_x\sin\theta-\hbar\lambda_y\cos\theta\right)\rho_x\sigma_x\no\\
&&+
\frac{v_Q\up\hbar}{2}\left[\left(J_x\cos\theta+\tau J_y\sin\theta\right)\rho_x\sigma_y-\left(J_x\sin\theta-\tau J_y\cos\theta\right)\rho_y\sigma_y\right]
+\tau\Delta\rho_z\sigma_x\,.
\eea

I have to remark that the above topological classification conclusions hold for a bulk system. In order to observe the two sets of edge MFs one must introduce boundaries
that preserve the ${\cal O}_u$ symmetry. The usual method followed in order to investigate the bulk-boundary correspondence of a translationally invariant topologically
non-trivial system, is to consider an infinite well potential. In the present case the preservation of ${\cal O}_u$ requires a specific behaviour under the
translation operation $t_{\pi/Q}$. As long as the approximation of Eq.~(\ref{eq:spinor expansion}) is well justified and an infinitely steep boundary potential is imposed, the
system is characterized by an emergent translational invariance and the two sets of edge MFs should manifestly appear. However, if the boundary potential $V_b(x)$ is not
infinetely steep but develops gradually within a certain region $l$, then the preservation of the ${\cal O}_u$ symmetry depends crucially on the the wave-vector $q_b\sim1/l$.
If $q_b$ is comparable to $q$, then the Fourier components $V_{q_b}$ have to be included in the bulk Hamiltonian of Eq.~(\ref{eq:qq}), contributing with terms proportional to
$\tau_z\rho_x$, $\rho_x$, $\tau_z\rho_y$ and $\rho_y$ that break the hidden symmetry. In this case, we may only observe only a single set of edge MFs. 

Nonetheless, the situation discussed here does not only constitute an example of mere academic interest, even if in the case where boundary effects can break the hidden
symmetry. Although the presence of protected boundary modes \cite{TI reviews,Boundary modes} is considered to be the hallmark of topologically non-trivial phases, it does not
constitute the unique route for diagnosing topological order. In fact, fingerprints of topological non-trivial phases can be also found in manifestations of the bulk system.
Consequently, we can obtain information concerning the presence of the ``hidden'' symmetry irrespective of the presence of boun\-daries. One example is the polar Kerr effect
\cite{PKE} that characterizes class D chiral p-wave superconductors. This experiment can provide a direct evidence of topological order by solely probing the bulk response.
Similar chiral phenomena emerge in non-superconducting systems. In the latter, apart from a similar polar Kerr effect \cite{Tewari PKE, PKE Graphene}, an anomalous
thermoelectric Nernst response \cite{Chirality Nernst,Niu Berry} and a topological Meissner effect \cite{Topological Meissner} constitute additional smoking gun signatures of
quantum anomalous Hall phases (class A). In fact, topological response survives also in finite temperatures, though exhibiting no quantization phenomena. Evenmore, the bulk
magnetic response \cite{Goryo magnetic response} of a quantum spin Hall insulator (class AII) can provide alternative routes for confirming the transition to the topologically
non-trivial phase.

\item \bt{Cases \RNum{3} and \RNum{4} with a single anti-unitary hidden symmetry} $\bm{{\cal O}}_a$

\noi Here I will investigate the consequences of the emergence of an anti-unitary hidden symmetry ${\cal O}_a$ on the symmetry class of the 1d model of Eq.~(\ref{eq:1d
model for magnetism and superconductivity}) that was obtained for Cases \RNum{3} and \RNum{4}. In this way I will be in a position to make a connection to previous studies
\cite{Flensberg spiral,Martin}. The Hamiltonian of Eq.~(\ref{eq:1d model for magnetism and superconductivity}) for $J_x=0$, has the following form
\bea
\widehat{{\cal
H}}(\hat{p}_x,x)=\left(\frac{\hat{p}_x^2}{2m}-\mu\right)\tau_z-\bm{M}(x)\cdot\left(\tau_z\sigma_x,\sigma_y,\tau_z\sigma_z\right)-\Delta\tau_y\sigma_y\,.
\label{eq:spin spiral}
\eea

First I will focus on the Case \RNum{3}, which belongs to class BDI with $\Theta={\cal K}$, if $\bm{M}(x)$ is random and $M_y(x)=0$. I demonstrate that the topological
properties of the system change with the emergence of a hidden symmetry. I now assume that $t_{\pi/Q}\bm{M}(x)=\bm{M}(x+\pi/Q)=-\bm{M}(x)$. For this special case, the
Hamiltonian is invariant under the anti-unitary hidden symmetry ${\cal O}_a=t_{\pi/Q}{\cal T}$. Since $t_{\pi/Q}$ is unitary and ${\cal T}$ anti-unitary, the particular
symmetry constitutes an additional generalized time-reversal symmetry $\widetilde{\Theta}={\cal O}_a$. In order to gain more insight, I will consider the simple spin-spiral
magnetization profile $M_x(x)=2M_x^Q\cos Qx$ and $M_z(x)=-2M_z^Q\sin Qx$ which is depicted in Fig.~\ref{fig:Spiral}. Prior studies \cite{Flensberg spiral,Martin} have focused
on the special case $M_x^Q=M_z^Q$. By expanding the field operator as in Eq.~(\ref{eq:spinor expansion}) we obtain
\bea
\widehat{{\cal H}}(\hat{p}_x)=\left[\frac{\hat{p}_x^2}{2m}+
\frac{(\hbar q)^2}{2m}-\mu\right]\tau_z+\tilde{v}\hat{p}_x\tau_z\rho_z-M_x^Q\tau_z\rho_x\sigma_x+M_z^Q\tau_z\rho_y\sigma_z-\Delta\tau_y\sigma_y\,.\label{eq:qqq}
\eea

\begin{figure}[t]
\includegraphics[width=4.5in,height=2.2in]{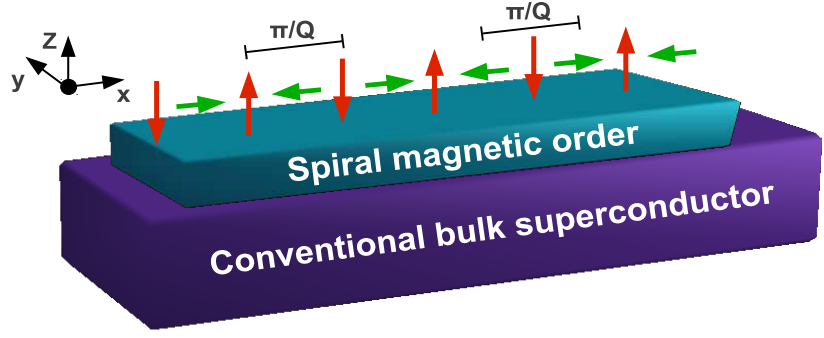}
\caption{Engineered {\color{black}TSC} consisting of a metal with spin-spiral magnetic order $\bm{M}(x)=\left(2M_x^Q\cos Qx,0,-2M_z^Q\sin Qx\right)$ placed on top of a bulk
conventional superconductor. The magnetic order is invariant under the combined action of time-reversal symmetry ${\cal T}$ followed by a $\pi/Q$ translation operation,
$t_{\pi/Q}$. The presence of the anti-unitary hidden symmetry ${\cal O}_a=t_{\pi/Q}{\cal T}$ leads to the class BDI$\oplus$BDI, compared to class BDI when it is
absent.}
\label{fig:Spiral}
\end{figure}

Within this framework we have $\Theta={\cal K}=\rho_x{\cal K}'$ and $\widetilde{\Theta}={\cal O}_a=t_{\pi/Q}{\cal T}=-i\rho_z i\sigma_y{\cal K}=
-i\rho_z i\sigma_y\rho_x{\cal K}'=i\rho_y\sigma_y{\cal K}'$. Remember that ${\cal K}'$ does not act on $q$ and $Q$. We readily observe that for both generalized
time-reversal symmetries we have $\Theta^2=\widetilde{\Theta}^2=+{\rm I}$ leading to the symmetry class BDI$\oplus$BDI. Note that in Refs.~\cite{Flensberg spiral,Martin} 
it was shown that, up to a spatially dependent unitary transformation, the model of Eq.~(\ref{eq:spin spiral}) is equivalent to the MF-wire model of Eq.~(\ref{eq:MF-wire})
with the latter belonging to the symmetry class BDI. However, as we showed here the particular system in the general case $M_x^Q\neq M_z^Q$ belongs to class BDI$\oplus$BDI.
It is straighforward to demonstrate that the two pictures agree with each other. By performing the transformation $\widehat{{\cal
H}}^{rot}(\hat{p}_x)=\Lambda^{\dag}\widehat{{\cal H}}^{rot}(\hat{p}_x)\Lambda$ with $\Lambda=(\rho_z\sigma_z+\sigma_y)(\rho_y+\rho_z)/2$ we obtain
\bea
\widehat{{\cal H}}^{rot}(\hat{p}_x)=\left[\frac{\hat{p}_x^2}{2m}+
\frac{(\hbar q)^2}{2m}-\mu\right]\tau_z+\tilde{v}\hat{p}_x\tau_z\rho_y+M_x^Q\tau_z\rho_z-M_z^Q\tau_z\rho_z\sigma_z-\Delta\tau_y\rho_y\sigma_z\,.
\eea

As anticipated, the above Hamiltonian is block diagonal and if we introduce the eigenstates $\sigma=\pm1$ of $\sigma_z$ in the rotated frame, we obtain
\bea
\widehat{{\cal H}}^{rot}_{\sigma}(\hat{p}_x)=\left[\frac{\hat{p}_x^2}{2m}+
\frac{(\hbar q)^2}{2m}-\mu\right]\tau_z+\tilde{v}\hat{p}_x\tau_z\rho_y+\left(M_x^Q-\sigma M_z^Q\right)\tau_z\rho_z-\sigma\Delta\tau_y\rho_y\,.
\eea

Remarkably, each of the above block Hamiltonians is identical to the MF-wire Hamiltonian of Eq.~(\ref{eq:MF-wire}) with effective chemical potential 
$\mu-(\hbar q)^2/2m$, spin-orbit coupling strength $\tilde{v}$, Zeeman field $M_x^Q-\sigma M_z^Q$ and superconducting order parameter $\sigma\Delta$. Each of the blocks
will be in the topologically non-trivial phase when the criterion 
\bea
\left|M_x^Q-\sigma M_z^Q\right|>\sqrt{\left[\mu-(\hbar q)^2/2m\right]^2+\Delta^2}\,,
\eea

is satisfied in complete analogy to Eq.~(\ref{eq:criterion}). We observe that the two sub-systems are not necessarily in the topologically non-trivial phase, at the same time.
Evenmore, if we consider $M_x^Q=M_z^Q$ as in prior studies \cite{Flensberg spiral,Martin}, only one of the sub-systems can be in the topologically non-trivial phase. In this
case, the system effectively behaves as a class BDI TSC and this is in accordance with the previous analytical findings. Note that as long as the translationally invariant
Hamiltonian is a good approximation and the boundary potential builds up spatially within a length much smaller that $1/q$, the hidden symmetry will be preserved and the
multiple edge MFs are expected to be observed when $M_x^Q\neq M_z^Q$.

\end{itemize}

\section{New TQC perspectives in a TSC with unitary discrete symmetries}

\noi For the cases that we considered in this work, hidden symmetry involved a specific behaviour under translations. However, this type of symmetry is fragile and can be
completely broken when boundaries are introduced. Nevertheless, one could look for alternative, robust and tunable, hidden symmetries that are related to some internal degree
of freedom such as a valley, orbital or band index. 

Let us now discuss new routes that open up for topological quantum computing when we consider the additional presence of a hidden unitary discrete symmetry ${\cal O}_u$.
Generally, two edge MFs $\gamma_a$ and $\gamma_b$ combine into a zero-energy fermion $d=(\gamma_a+i\gamma_b)/\sqrt{2}$ that leads to a doubly degenerate ground state
$\left|1\right>$ and $\left|0\right>$. These two states correspond to many-body ground states where the zero-energy fermion is occupied or not, respectively. The Hilbert space
spanned by these two degenerate states defines a topological qubit which is in principle \cite{MFdeco} protected by decoherence due to the non-local binding of the MFs. The
fundamental and in fact the only allowed topologically protected single qubit operation which we may perform is called braiding \cite{TQC,Alicea} and corresponds to exchanging
the two MFs in real space Fig.~\ref{fig:Braiding}. After braiding is effected the states $\left|1\right>$ and $\left|0\right>$ become $e^{i\pi/4}\left|1\right>$ and
$e^{-i\pi/4}\left|0\right>$, picking up a relative $\pi/2$ phase. For a counterclockwise rotation, braiding corresponds to the transformation $\gamma_a\rightarrow+\gamma_b$
and $\gamma_b\rightarrow-\gamma_a$ while for a clockwise rotation we obtain the inverse transformation $\gamma_a\rightarrow-\gamma_b$ and $\gamma_b\rightarrow+\gamma_a$
\cite{Alicea}.

\begin{figure}[h]
\includegraphics[width=0.2\textwidth]{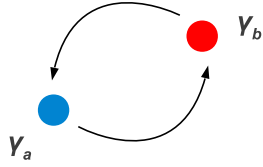}
\caption{Exchanging Majorana fermions in real space leads to a protected single topological qubit rotation termed braiding. In qubit space, this corresponds to a gate 
operation where the qubit states pick up a relative $\pi/2$ phase.}
\label{fig:Braiding}
\end{figure}

For a system that supports a single MF per edge, the emergence of a unitary discrete symmetry ${\cal O}_u$ with ${\cal O}_u^2={\rm I}$, can lead to an additional MF per
edge. The two MFs per edge are labelled by the $\pm1$ eigenstates of ${\cal O}_u$, leading to the following 4 edge MFs $\gamma_{a\pm}$ and $\gamma_{b\pm}$. With the latter, we
can define two zero-energy fermions $d_{\pm}=(\gamma_{a\pm}+i\gamma_{b\pm})/\sqrt{2}$ and two topological qubits with states $\{\left|1_+\right>,\left|0_+\right>\}$ and
$\{\left|1_-\right>,\left|0_{-}\right>\}$. The accessible protected non-Abelian operations that we may perform within this four-fold degenerate Hilbert space are
restricted by the \textit{simultaneous conservation of fermion parity and ${\cal O}_u$}. Essentially, the possible operations are combinations of simultaneous or separate
clockwise and counterclockwise braiding operations in each of the topological qubit spaces. Specifically, we have the following four operations:
\bea
A:\{\left|1_+\right>,\left|0_+\right>,\left|1_-\right>,\left|0_{-}\right>\}&\rightarrow&
\{e^{i\pi/4}\left|1_+\right>,e^{-i\pi/4}\left|0_+\right>,\left|1_-\right>,\left|0_{-}\right>\}\,,\\
B:\{\left|1_+\right>,\left|0_+\right>,\left|1_-\right>,\left|0_{-}\right>\}&\rightarrow&
\{\left|1_+\right>,\left|0_+\right>,e^{i\pi/4}\left|1_-\right>,e^{-i\pi/4}\left|0_{-}\right>\}\,,\\
C:\{\left|1_+\right>,\left|0_+\right>,\left|1_-\right>,\left|0_{-}\right>\}&\rightarrow&
\{e^{i\pi/4}\left|1_+\right>,e^{-i\pi/4}\left|0_+\right>,e^{i\pi/4}\left|1_-\right>,e^{-i\pi/4}\left|0_{-}\right>\}\,,\\
D:\{\left|1_+\right>,\left|0_+\right>,\left|1_-\right>,\left|0_{-}\right>\}&\rightarrow&
\{e^{i\pi/4}\left|1_+\right>,e^{-i\pi/4}\left|0_+\right>,e^{-i\pi/4}\left|1_-\right>,e^{+i\pi/4}\left|0_{-}\right>\}\,,
\eea

\begin{figure}[b]
\includegraphics[width=0.17\textwidth]{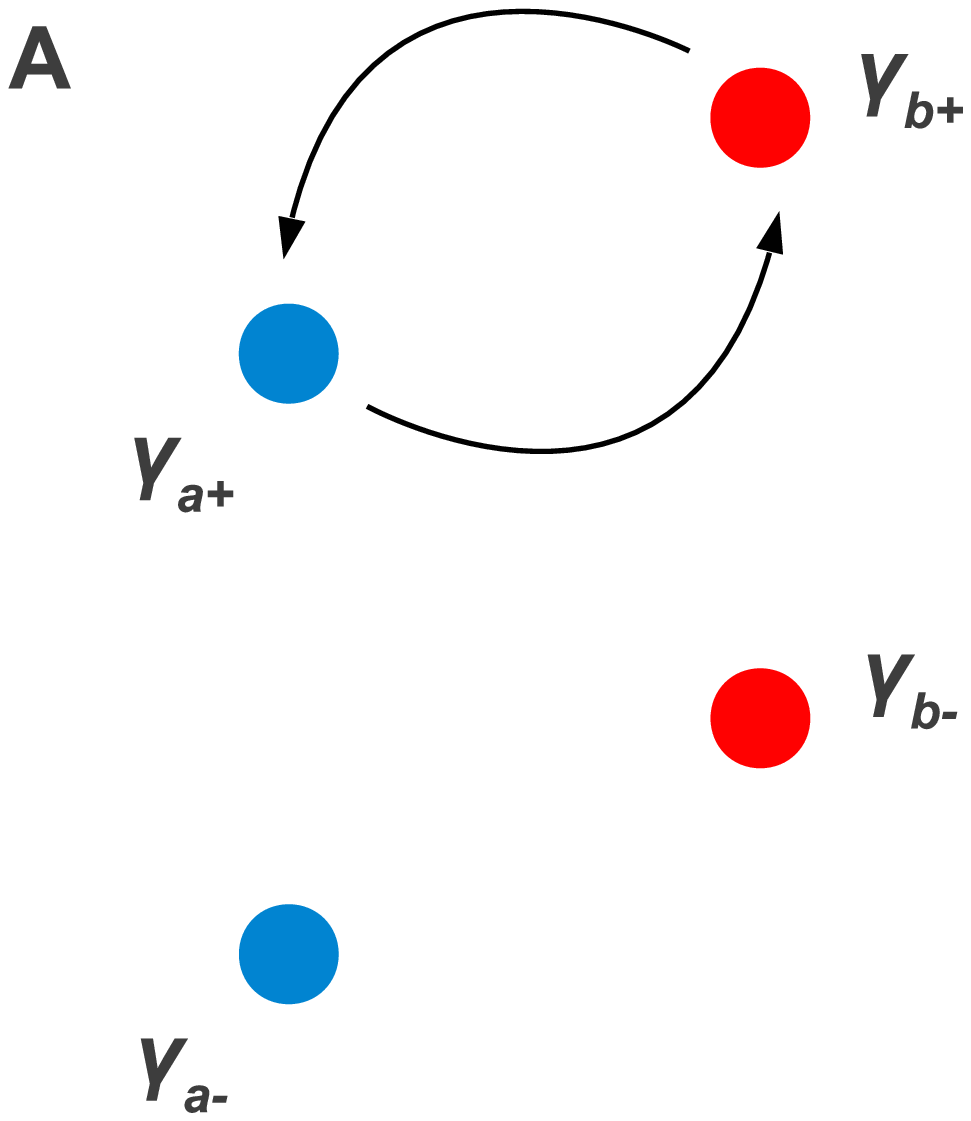}\hspace{0.6in}\includegraphics[width=0.17\textwidth]{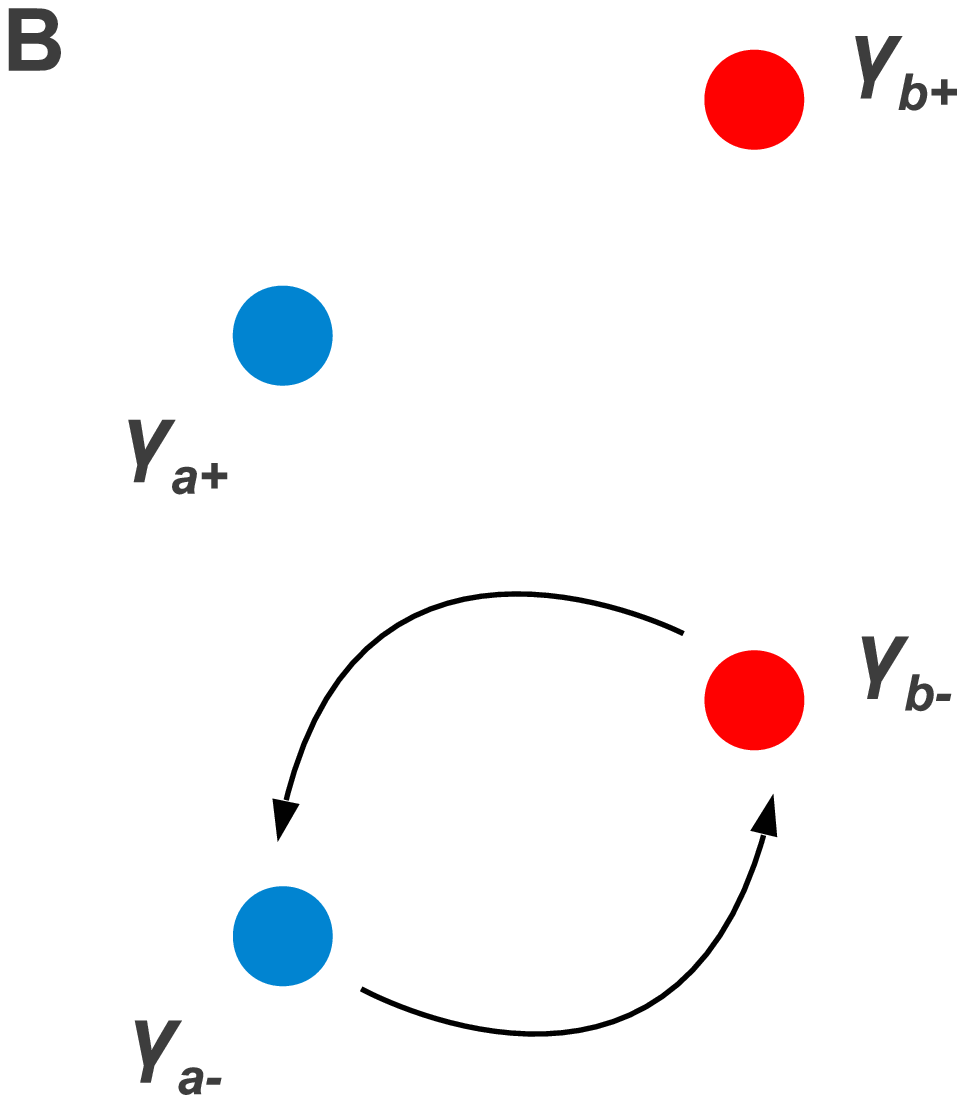}\hspace{0.6in}
\includegraphics[width=0.17\textwidth]{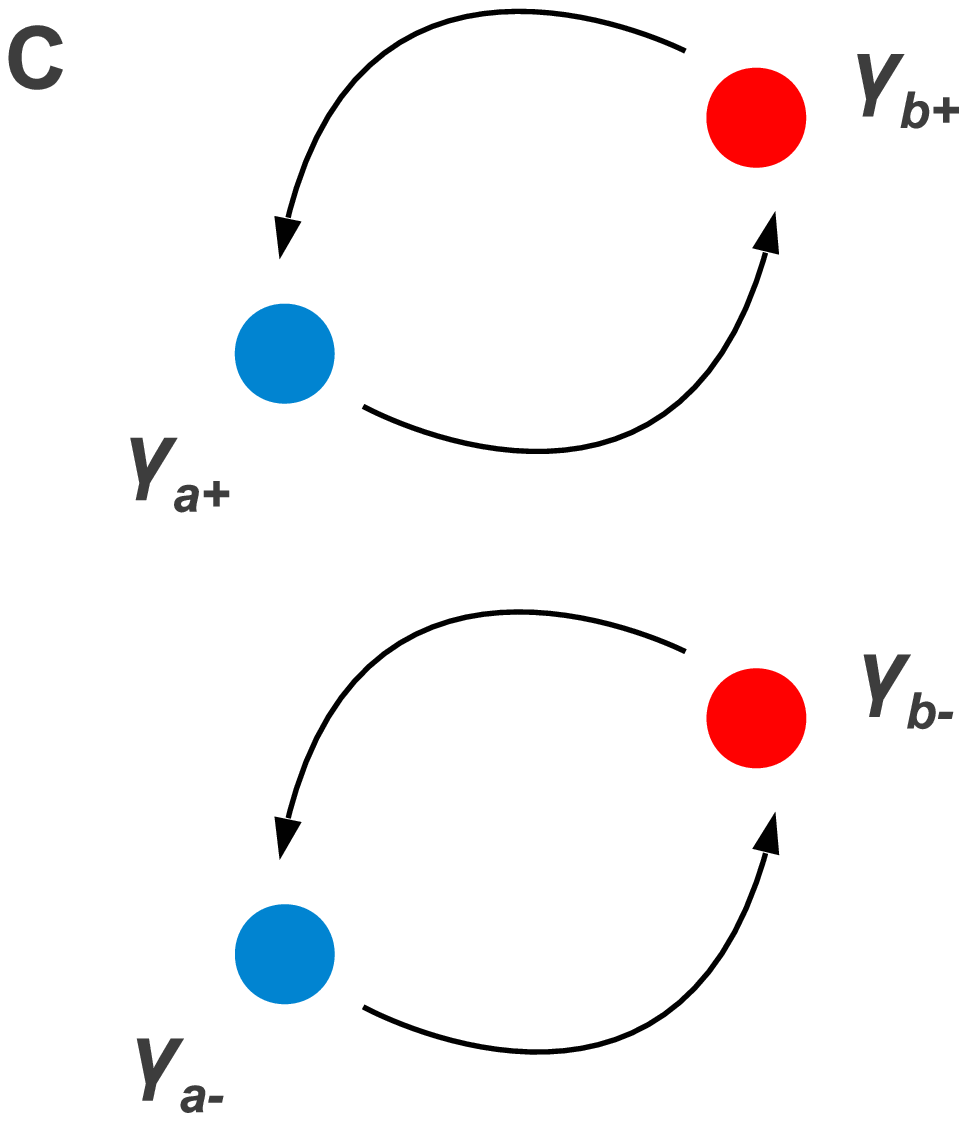}\hspace{0.6in}\includegraphics[width=0.17\textwidth]{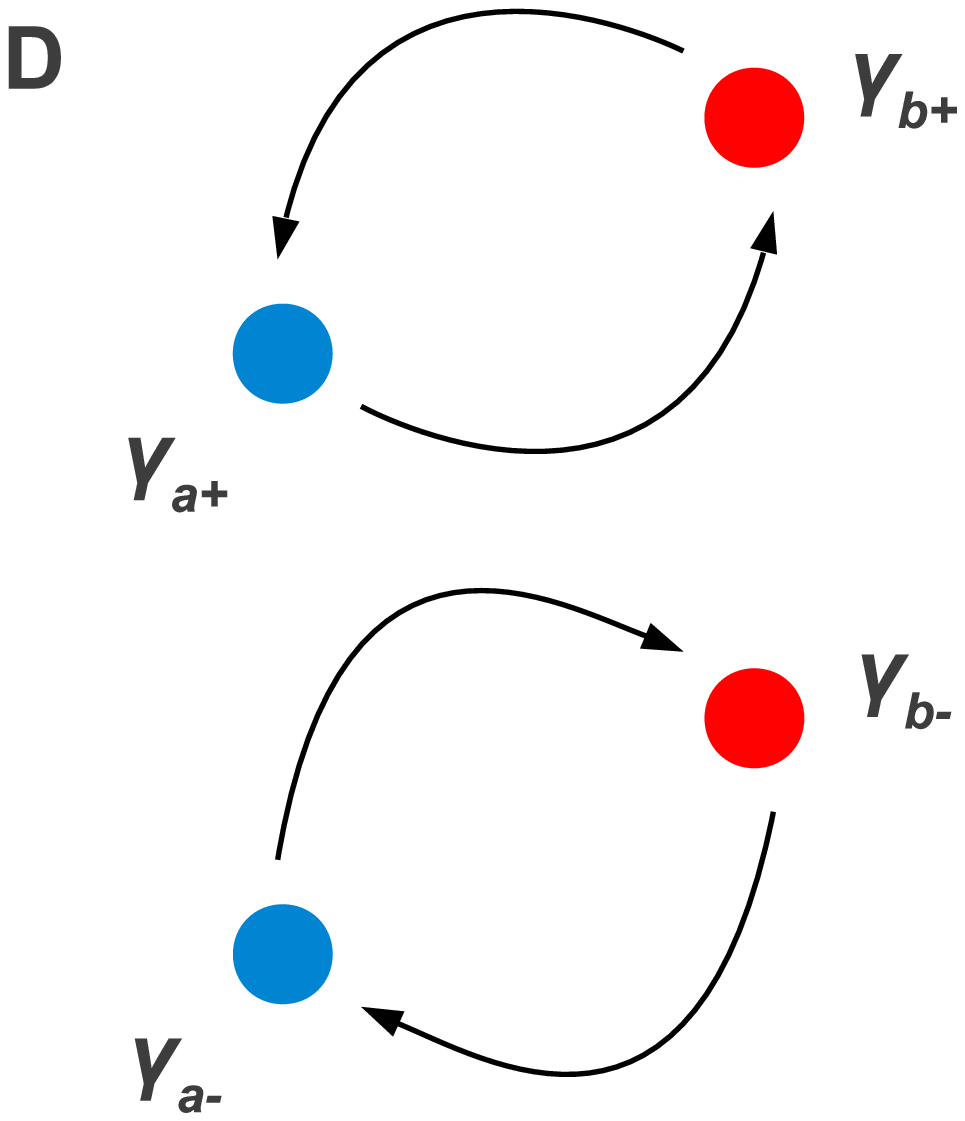}
\caption{Allowed topologically protected operations for two topological qubits $\pm$ corresponding to the $\pm 1$ eigenvalues of a unitary hidden symmetry operator ${\cal
O}_u$ satisfying ${\cal O}_u^2={\rm I}$. The standard braiding operations A and B describe single qubit operations. C and D correspond to operations in the joint qubit space.
In operation C, the $\pm$ pairs of Majorana fermions are braided in the same direction while in case D in the opposite.}
\label{fig:Braiding2qubits}
\end{figure}

presented in Fig. \ref{fig:Braiding2qubits}. The operations $A$ and $B$ correspond to braiding operations effected only on the $+$ or the $-$ topological qubits. These are
single qubit operations. In contrast, if we effect braiding simultaneously in both qubits spaces we have two options. Either the direction of braiding is the same or opposite.
These two possible topologically protected operations in the joint qubit space are described by the $C$ and $D$ configurations. The above set of protected operations do not
suffice for performing universal TQC due to the Ising nature of the MFs \cite{TQC,UTQC}. Nevertheless, the presence of the additional topological qubit on the \textit{same}
wire can be useful for performing braiding operations. So far, several methods for performing braiding have been proposed, including networks of topological wires \cite{Fu
and Kane,braiding with networks} where neighbouring MFs can be controllably coupled in order to perform a MF exchange. In the present case, the additional ${\cal O}_u$
protected MFs can constitute a reservoir of MFs that could reduce the number of complementary wires that one needs for performing adiabatic operations using these protocols.
In addition, the presence of the extra pair of MFs can be also prominent for creating phase gate operations. A standard theoretical proposal \cite{TQC} for implementing a
phase gate for two separated MFs, prescribes to bring the MFs to a finite distance in order to let them hybridize into a finite energy fermionic state. Due to the time
evolution of the finite energy state, a phase gate operation will be implemented on the topological qubit when the MFs reseparate. In the presence of a hidden symmetry ${\cal
O}_u$, one does not have to change the distance of the MFs any more. By controllably switching off the hidden symmetry ${\cal O}_u$, one hybridizes the two MFs of the same
edge, for instance $\gamma_{a\pm}$, so to end up with a single MF. Depending on the details of the hidden symmetry breaking and restoration procedures, one may retrieve a
phase gate operation. Of course a detailed investigation of these possibilties is required.

The alternative TQC routes described above depend delicately and crucially on the robustness of this hidden symmetry. As we have already mentioned, it is desirable to find a
system that has a hidden symmetry related to a degree of freedom such as a band index. For example, in the case of a two-band topological superconductor where only intraband
matrix elements appear in the Hamiltonian, the system splits into two irreducible subsystems similarly to the situation described above. As a matter of fact, multiband systems
such as the Fe-based high-Tc superconductors \cite{Basic FeAs}, offer a promising way out. The latter materials are supposed to exhibit intra-band superconductivity (usually a
4-band \cite{FeAs 4band} or a 5-orbital \cite{FeAs 5orbital} picture is adequate) and consequently we may obtain a number of disconnected sub-systems. If we manage to render
each of these superconducting sub-systems topological, we will be in a position to apply the topological quantum computing protocols discussed in the previous paragraph.
Recently, a proposal concerning topological superconductivity based on iron-based superconductors has been put forward \cite{Kane FeAs}. However, in that work an iron-based
superconductor was used to induce superconductivity by proximity effects on a Rashba-semiconductor. Instead, the situation that I envisage involves intrinsic multiband
topological superconductivity in the iron pnictide superconductor itself.

\section{Conclusion}

I have performed a detailed analysis of the accessible topological superconducting phases that can occur from the combination of inhomogeneous Rashba spin-orbit coupling,
magnetic order and superconductivity. By exploring the landscape of the possible topological phases I proposed new systems prominent for realizing MFs, based on Rashba
spin-orbit coupling and ${\cal T}$ violating superconductivity, without the demand for any kind of magnetic order. Specifically, I explicitly demonstrated the emergence of
MFs in a platform consisting of two coupled single channel Rashba semiconducting wires deposited on top of a Josephson junction fabricated by conventional superconductors.
Moreover, I pinpointed the significance of emergent unitary and anti-unitary hidden symmetries and revealed the topological implications that they lead to. Finally, I
discussed alternative topological quantum computing pathways that open up in the presence of a unitary hidden symmetry and suggested that Fe-based multiband superconductors
could be a potential candidate for these implementations. 

\section*{Acknowledgements}

I am grateful to Gerd Sch\"{o}n, Alexander Shnirman, Georgios Varelogiannis, Jens Michelsen and Daniel Mendler for the motivation and the great support. Furthemore, I am also
indebted to Xiao-Liang Qi, Ivar Martin, Alberto Morpurgo, Andreas Heimes, Stefanos Kourtis, Lingzhen Guo, Juan Atalaya, Boris Narozhny, Mathias Scheurer, Bhilahari Jeevanesan,
Elio K\"{o}nig, Michael Marthaler and Paris Parisiades for numerous suggestions and enlightening discussions. In addition, I acknowledge financial support from the EU project
NanoCTM.


\begin{thebibliography}{00}
\bibitem{commentU} Here, the characterization ``unconventional'' is assigned to order parameters of density wave or superconducting phases, that carry finite angular  momentum
such as p-wave, d-wave, etc.
\bibitem{comment} {\color{black}The 2d and quasi-1d systems always belong to the same symmetry class. Here due to some specific spatial dependences of certain parameters, the
symmetry of the quasi-1d system can be enhanced, leading to a different symmetry class. However, this is an artifact of the choices made here for simplifying the discussion.
It is assumed that we may always add terms in the quasi-1d Hamiltonian which ensures that all the possible unitary symmetries are broken.}
\end{thebibliography}
\end{document}